\documentclass[fleqn, usenatbib]{mnras}

\usepackage{newtxtext,newtxmath}
\usepackage[T1]{fontenc}
\usepackage{ae,aecompl}
\usepackage{ulem}

\usepackage{graphicx}	
\usepackage{amsmath}	

\newcommand{\pluto}{\textsc{pluto }}
\newcommand{\cloudy}{\textsc{cloudy }}

\newcommand{\psec}{\mbox{ s}^{-1} }

\newcommand{\pcc}{\mbox{ cm}^{-3} }
\newcommand{\pcm}{\mbox{ cm}^{-1} }
\newcommand{\pcmsq}{\mbox{ cm}^{-2} }
\newcommand{\msun}{\mbox{ M}_{\odot}}
\newcommand{\mpy}{\mbox{ M}_{\odot} \mbox{yr}^{-1} }
\newcommand{\ergps}{\mbox{ erg s}^{-1} }
\newcommand{\ergpspcmsq}{\mbox{ erg s}^{-1} \mbox{cm}^{-2} }
\newcommand{\kmps}{\mbox{km s}^{-1}}
\newcommand{\tcool}{t_{\rm cool} }
\newcommand{\tcoolon}{t_{\rm cool, onset} }
\newcommand{\tcoolend}{t_{\rm cool, end} }
\newcommand{\rbub}{r_{\rm bub}}
\newcommand{\cie}{\textit{CIE}}
\newcommand{\nei}{\textit{NEI}}

\newcommand{\nsrd}{\textit{NSRD}}
\newcommand{\nsrdc}{\textit{NSRDC}}

\newcommand{\HI}{H$^0$ }
\newcommand{\HII}{H$^{+}$ }
\newcommand{\HeI}{H$^{0}$ }
\newcommand{\HeII}{He$^{+}$ }
\newcommand{\nii}{N$^+$ }
\newcommand{\niii}{N$^{2+}$ }
\newcommand{\niv}{N$^{3+}$ }
\newcommand{\nv}{N$^{4+}$ }
\newcommand{\nvi}{N$^{5+}$ }
\newcommand{\nvii}{N$^{6+}$ }
\newcommand{\cii}{N$^{+}$ }
\newcommand{\ciii}{N$^{2+}$ }
\newcommand{\civ}{C$^{3+}$ }
\newcommand{\oii}{O$^+$ }
\newcommand{\oiii}{O$^{2+}$ }
\newcommand{\ovi}{O$^{5+}$ }
\newcommand{\ovii}{O$^{6+}$ }
\newcommand{\oviii}{O$^{7+}$ }
\newcommand{\sii}{S$^+$}
\newcommand{\xnii}{x_{N^+}}
\newcommand{\nnii}{N_{N^+}}
\newcommand{\xniii}{x_{N^{2+}}}
\newcommand{\nniii}{N_{N^{2+}}}
\newcommand{\xnv}{x_{N^{4+}}}
\newcommand{\nnv}{N_{N^{4+}}}
\newcommand{\rs}{r_{\rm shock}}

\title[Evolution of SN]{Non-equilibrium ionisation and radiative transport in an evolving supernova remnant}

\author[Sarkar, Gnat \& Sternberg]{
Kartick C. Sarkar,$^{1}$\thanks{E-mail: sarkar.kartick@mail.huji.ac.il, kartick.c.sarkar100@gmail.com}
and Orly Gnat $^{1}$
Amiel Sternberg$^{2,3,4}$\\
$^{1}$Racah Institute of Physics, The Hebrew University of Jerusalem, 91904, Israel\\
$^{2}$School of Physics and Astronomy, Tel Aviv University, Ramat Aviv, 69978, Israel\\
$^{3}$Centre for Computational Astrophysics, Flatiron Institute, 162 5th Avenue, 10010, New York, NY, USA\\
$^{4}$Max-Planck-Institut fur Extraterrestrische Physik (MPE), Giessenbachstr., 85748, Garching, FRG
}

\date{Accepted XXX. Received YYY; in original form ZZZ}
\begin{document}

\label{firstpage}
\pagerange{\pageref{firstpage}--\pageref{lastpage}}
\maketitle

\begin{abstract}
We present numerical simulations of the evolution of a supernova (SN) remnant expanding into a uniform background medium with density $n_H = 1.0\,\pcc$  and temperature of $10^4$ K.  We include a dynamically evolving non-equilibrium ionisation (NEI) network (consisting of all the ions of H, He, C, N, O, Ne, Mg, Si, S, Fe), frequency dependent radiation transfer (RT), thermal conduction, and a simple dust evolution model, all intra-coupled to each other and to the hydrodynamics. We assume spherical symmetry. Photo-ionisation, radiation losses, photo-heating, charge-exchange heating/cooling and radiation pressure are calculated on-the-fly depending on the local radiation field and ion fractions. We find that the dynamics and energetics (but not the emission spectra) of the SN remnants can be well modelled by collisional equilibrium cooling curves even in the absence of non-equilibrium cooling and radiative transport. 
We find that the effect of precursor ionising radiation at different stages of SN remnant are dominated by rapid cooling of the shock and differ from steady state shocks. The predicted column densities of different ions such as \nii, \civ and \nv, can be higher by up to several orders of magnitude compared to steady state shocks. We also present some high resolution emission spectra that can be compared with the observed remnants to obtain important information about the physical and chemical states of the remnant, as well as constrain the background ISM.
\end{abstract}

\begin{keywords}
ISM:supernova remnants, HII regions, bubbles -- methods:numerical -- radiative transfer, hydrodynamics
\end{keywords}


\section{Introduction}
\label{sec:intro}
Radiative and mechanical feedback processes from supernovae (SNe) are critical for the evolution of galaxies, influencing small scale structures and phases states in the interstellar medium (ISM), to global star-formation regulation in galaxy evolution across cosmic time \citep{Larson1974, McKee1977, Dekel1986, Nath1997, Cox2005, Breitschwerdt2006, Krumholz2018, Dekel2019}.
Study of individual and clustered SNe is, therefore central for understanding injections of mass, momentum, energy and metals at $\sim 10$ pc scales, not readily resolved in larger scale galaxy formation simulations.

Apart from injecting energy, momentum and metals to the ISM, SN shocks are also important sources of many optical/IR/UV/X-ray emission lines observed in galaxies.
A detailed understanding of the emission lines produced in SN remnants (SNR) is required to distinguish between pure photo-ionised and shock excited line emissions galactic ISM patches. 
Most of the existing distinctions are based on different optical line ratios, like [\ion{Si}{ii}]/H$\alpha$, [\ion{O}{i}]/H$\alpha$, [\ion{O}{iii}]/H$\alpha$ etc \citep{Mathewson1973, Fesen1985, Kewley2001, Kopsacheili2020} where shocks are modelled assuming a steady state rather than a complete time-dependent evolution. Optical lines from individual SNR have also been used to infer the metallicity gradient in an external galaxy or in our Galaxy based on such steady state models \citep{Mathewson1973, Dopita1976, Dopita1980, Fesen1985, Dopita2019}.

The evolution of individual radiative SN remnants has been studied extensively in the literature \citep{McKee1977, Cioffi1988, Slavin1992, Kim2015, Steinwandel2020} . Although the importance of non-equilibrium ionisation and cooling was initially neglected, later studies included such complex physics. \cite{Hamilton1983, Kafatos1973, Gnat2007} studied the NEI evolution of different ions for  
a time-dependent temperature history.
Given the temperature history of each cells/particles behind the supernovae shocks one can calculate the full ionisation dynamics. Self consistent numerical simulations with non-equilibrium ionisation networks have also been studied to infer abundances of ions such as \ovi, \nv and Si$^{3+}$ for comparison to observations in the local ISM \citep{Slavin1992, DeAvillez2012}. 

While radiation loss from the SN
is important in setting its dynamics, its effect on the interstellar medium and the SN itself has not been studied with proper geometry and time evolution. Attempts have been made to incorporate the radiative transfer along with ionisation network. Such attempts however, remained limited to only steady state shocks \cite{Shull1979,Dopita1996,Gnat2009, Sutherland2017} that do not consider either the geometrical factors or the full structure of a SN bubble/remnant. Recently \cite{Zhang2018, Zhang2019, Steinwandel2020} studied non-equilibrium 
chemistry in a SN remnant but they did not include any self-radiation from the SN remnant. In order to overcome such limitations, we, for the first time, study the full evolution of a SN structure in 
a uniform background medium including self consistent ionisation network, radiative transfer, conduction and a simple evolution of dust.

Our paper is organised as follows. We describe our numerical tools and simulation details in section \ref{sec:simulation-method}. We present our results for the dynamics and evolution of SN shock in section \ref{sec:dynamics}. In section \ref{sec:observables}, we present our estimations of different column densities and compare them with traditionally used steady state shock models. A brief discussion of the limitations of our current work is described in section \ref{sec:discussion}. A summary presented in section \ref{sec:summary}.

\section{Simulation method} 
\label{sec:simulation-method}
We perform spherically symmetric simulations using a non-equilibrium ionisation (NEI) network plus radiative transfer (RT) module, described and tested in \cite{Sarkar2020a} (hereafter, paper-I), based on our updated version of the magnetohydrodynamics (MHD) code \pluto \citep{Mignone2007}. We provide a brief description of our code and the initial setup in the following sections. We refer the reader to \cite{Mignone2007, Tesileanu2008} and paper-I for further technical details.

\subsection{The code}
\label{subsec:the-code}
 \pluto is an Eulerian grid code 
 that uses the finite volume method to solve the fluid equations. The source terms are solved by operator splitting. The full set of hydrodynamics equations are 
\begin{eqnarray}
&&\frac{\partial}{\partial t} \rho  + \frac{1}{r^2} \frac{\partial}{\partial r}\left(r^2 \rho v\right) = \dot{\rho}_s \\
\label{eq:rho-cons}
&&\frac{\partial}{\partial t} \left(\rho v\right) + \frac{1}{r^2} \frac{\partial}{\partial r}\left(r^2\rho v^2\right) = -\frac{\partial}{\partial r} p + \rho a_r \\
\label{eq:mom-cons}
&&\frac{\partial}{\partial t} E + \frac{1}{r^2} \frac{\partial}{\partial r}\left[r^2 v\left(E + p \right) \right] \nonumber \\
&& \quad\quad\quad\quad\quad= \mathcal{H}-\mathcal{L}  + \rho\: v\: a_r + \frac{1}{r^2} \frac{\partial}{\partial r} \left( r^2 F_c\right)\\
\label{eq:energy-cons}
&& \frac{\partial}{\partial t} \left(\rho X_{k,i}\right) + \frac{1}{r^2} \frac{\partial}{\partial r}\left(r^2 v\:\rho X_{k,i} \right) = \rho S_{k,i} 
\label{eq:IN}
\end{eqnarray}
where $\rho\equiv \rho(r)$ is the density, $v \equiv v(r)$ is the velocity, $p \equiv p(r)$ is thermal pressure, $E = p/(\gamma-1) + \rho v^2/2$ is the total energy density and $\gamma=5/3$ is the adiabatic index. The source terms, $\dot{\rho}_s$, $\mathcal{H}$ and $\mathcal{L}$ represent the mass injection rate, thermal heating (via photo-ionisation and charge exchange) and thermal cooling (radiative emissions, recombinations and charge exchange), respectively. The effect of radiation force on momentum and energy is incorporated by the radiation acceleration term $a_r$. The conductive flux is given by $F_c = F_{\rm sat}/(F_{\rm sat} + |F_{\rm class}|)\times F_{\rm class}$, where $F_{\rm class} = 5.6\times 10^{-7}\: T^{5/2}\: \partial T/\partial r\: \ergpspcmsq $ \citep{Spitzer1956} and $F_{\rm sat} = 5 \phi\:\rho\:c_{\rm iso}^3$ with $\phi = 0.3$ \citep{Cowie1977}.

The ionisation network is incorporated through Eqn \ref{eq:IN}, where $X_{k,i}$ is the individual ion fraction of element $k$ in its $i$-th ionisation level. The abundances of the elements are kept constant and equal to the Solar values throughout our simulations. The source function, $S_{k,i} = S(X_{k,i-1}, X_{k,i}, X_{k, i+1}, \psi(\mu, \nu))$ represents the ionisation/recombination/photo-ionisation/Auger rates for an ion $(k,i)$ and depends on the local radiation spectra $\psi(\mu=\cos\theta, \nu)$.  This equation, therefore, is a set of $111$ coupled ODEs. The instantaneous cooling ($\mathcal{L}$) and heating ($\mathcal{H}$) functions are calculated based on the local NEq ion fractions and radiation spectra. 

To obtain the local radiation spectra, we solve for the direction dependent specific intensity $\psi(\mu, r, \nu)$ in a spherically symmetric system by assuming that the change in hydrodynamic or chemical properties in a cell occurs much more slowly than the light travel time across the system and  that the scattering is negligible compared to absorption \footnote{This is true for most of the frequency range we consider for our RT. One exception is the resonant scattering of Ly$\alpha$ line which requires a better transport solver than presented here. Although the Ly$\alpha$ scattering is, in principle, taken care of by the local emissivity to some extent, the effect only takes place at the next step rather than instantaneously.}. The RT equation is then
\begin{eqnarray}
\frac{\mu}{r^2}\, \frac{\partial}{\partial r}\left( r^2\, \psi(\mu, r,\nu)\right) &+& \frac{1}{r} \frac{\partial}{\partial \mu}\left( (1 - \mu^2) \psi(\mu, r,\nu)\right) \nonumber \\
&=& j_\nu - \alpha_\nu\, \psi(\mu, r,\nu) \,,
\label{eq:RTE-sph}
\end{eqnarray}
where, $\mu =\cos\theta$ (with $\theta$ being the angle between a ray and radial direction), $j_\nu$ is the isotropic emissivity, $\alpha_\nu$  is the absorption coefficient ($\pcm$, which we loosely refer to as \textit{opacity}). Notice that although our system is a 1D spherical system, the solution of a frequency dependent RT equation is a 2D axisymmetric problem. 

We also include a simple dust prescription in our simulations. The dust provides extinction (absorption+scattering) and experiences radiation pressure which is assumed to couple instantaneously to the gas. The initial dust properties (extinction cross section, albedo and average scattering angle) are assumed to be frequency dependent 
as described by \cite{Weingartner2001} for $R_v = 3.1$ which is close to the dust properties observed in Milky-Way\footnote{Also available in \url{https://www.astro.princeton.edu/~draine/dust/dustmix.html}}. The extinction cross section decreases when the dust is subjected to thermal sputtering \cite{Draine2011}. We assume that the suppression is proportional to $a^2$, where $a = 0.1 \mu$m is the initial size of a typical dust grain. Although there can be other processes like shattering and evaporation due to shock propagation, the strength of these processes depends on the shock speed which is also represented by the shock temperature. Our dust destruction prescription, therefore, is very simple and works only to estimate a relative change in the extinction.

A detailed discussion of the techniques and frequencies used can be found in paper-I.

\begin{table}
 \caption{List of simulations performed in this paper.}
 \label{table:list-of-runs}
 \setlength{\tabcolsep}{4pt}
 \begin{tabular}{lcccccc}
 \hline\hline
 Name & NEI & Self- & Dust & Conduction & Density & resolution\\
  & & radiation & & & [H/cm$^3$] & [pc]$^{**}$ \\
 \hline
 CIE$^{*}$ & No & No & No & No & 1.0 & 0.001 \\
 NEI & Yes & No & No & No & 1.0 & 0.006, 0.001 \\
 NSR & Yes & Yes & No & No & 1.0 & 0.006, 0.001 \\
 NSRD & Yes & Yes & Yes & No & 1.0 & 0.006, 0.001 \\
 NSRDC & Yes & Yes & Yes & Yes & 1.0 & 0.006, 0.001 \\
 \hline
 \end{tabular}
  {\\$^*$ In this case we assume a plasma in collisional equilibrium.\\ $^{**}$In the cases where two resolutions are mentioned, we apply the lower resolution at $r = 0.1-18$ pc and the higher resolution at $r = 18-50$ pc to resolve the shell better and to lower computational cost. The resolution mentioned here is the default value. We vary this value while checking for convergence with resolution.}
 \end{table}

\subsection{Initial and boundary condition}
\label{subsec:initial-cond}
Initially we set the box to have uniform density with hydrogen number density, $n_{0}$ at a temperature of $T_{\rm amb} = 10^4$~K and Solar metallicity.  We allow the gas to radiatively cool to a floor temperature of $6\times 10^3$ K. We set the initial ionisation state of the medium to collisional equilibrium at $T = T_{\rm amb}$. The simulation box extends from $r= 0.1$ pc to $50$ pc.

The SN energy is injected by placing $5 \msun$ of gas with $10^{51}$ erg of internal energy within a radius of $1$ pc at $t = 0$. The inner and outer boundary conditions for the hydrodynamic and chemical quantities are set to an outflow condition, i.e. copied from the inner cells to the ghost zones (no gradient across the boundary). 
The incoming ($\mu<0$) radiation spectrum at the outer boundary is assumed to be a uniform \cite{Haardt2012} background  \footnote{The background radiation does not penetrate more than a few pc from the outer surface due to high absorption cross section by the neutral hydrogen at our considered densities} for redshift $z=0$. The inner boundary condition for the radiation spectrum is set to be reflective i.e. $\psi(\mu\leq 0, \nu) = \psi(\mu\geq 0, \nu)$. This is possible due to the spherical symmetry of the problem.

We perform different simulations with increasing complexity for $n_0 = 1.0 \pcc$. 
First, we assume pure collisional equilibrium for the ionisation states of the gas (case, \cie). 
Next, we include the non-equilibrium ionisation network to calculate the ionisation states of the gas on-the-fly (case \nei). We include our calculation of self-radiation and couple it to the ionisation network in case \nsrd. This case also includes extinction and scattering from dust. 
Finally, we add thermal conduction along with the ionisation network and self-radiation in run \nsrdc. 
Although the \nsrd~and \nsrdc ~models include a simple prescription of dust evolution, we do not expect the dust to play any major role at $n_0 \lesssim 100 \pcc$ (see paper-I).

Within our $0.1-50$~pc simulation box, we apply two types of uniform spatial resolution 
to save computation time. A lower resolution is applied for $r=0.1-18$~pc where the shock is still self-similar ($t\lesssim 23$ kyr) and has not undergone much cooling. 
A higher resolution is applied for $r=18-50$~pc, where the shock undergoes rapid cooling and shell formation, and is prone to resolution effects. For the case with conduction (\nsrdc), we apply the high resolution grids only up to $40$ pc (since by $300$~kyr, the shock only reaches this point).
The grid from $40-50$ pc is set to have the lower resolution. We sample the angular direction, $\mu$,  with $16$ uniformly spaced rays between $-1$ to $+1$ for the purpose of the radiative transfer.

A full list of runs can be found in Table \ref{table:list-of-runs}. The labels indicate the physics each run includes. 


\begin{figure*}
	\centering
	\includegraphics[width=0.75\textwidth, clip=true, trim={0cm 0.5cm 0cm 0.5cm}]{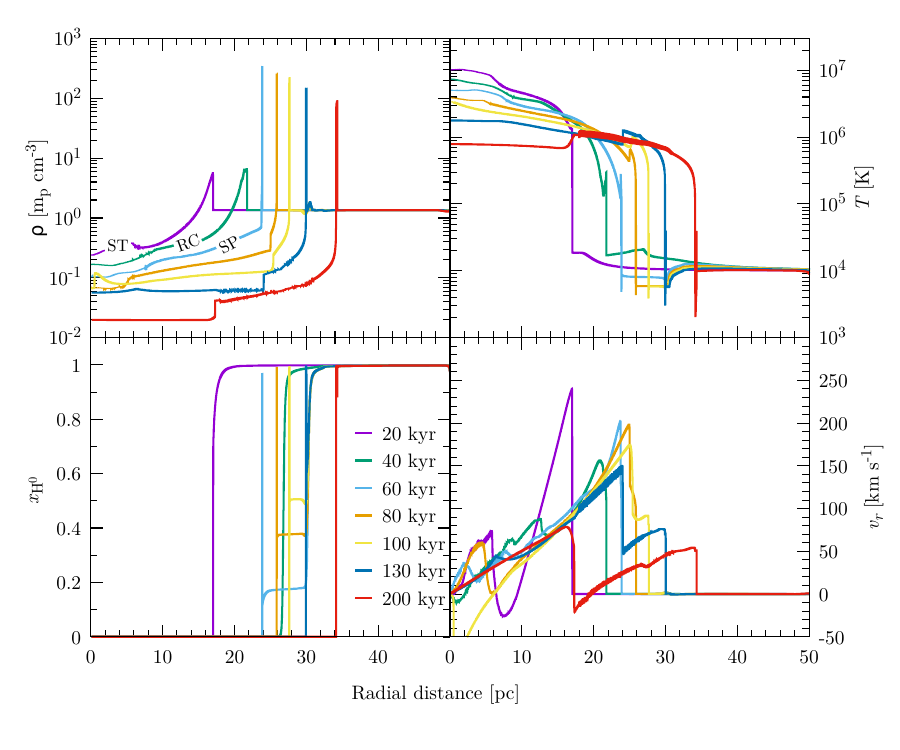}
	\caption{Evolution of density ($\rho$, top left), temperature ($T$, top right), Hydrogen ionisation fraction ($x_{\rm H^0}$, bottom left) and radial velocity ($v_r$, bottom right) structure for the 
	\nsrdc~case (see table~1). Typical phases of the SN can be seen clearly. The structure is in Sedov-Taylor (ST) phase at $20$ kyr, in rapid cooling (RC) phase at $40$ kyr and in snow-plow (SP) phase at $t > 60$ kyr. The little bump in front of the shell at $t = 130$ kyr is the stationary ionisation front for all the previous phases. The shell reaches the static ionisation front at $t \sim 130$ kyr.}
	\label{fig:rho-T-x-vr}
\end{figure*}

\begin{figure}
	\centering
	\includegraphics[width=0.48\textwidth, clip=true, trim={0cm 0cm 0cm 0cm}]{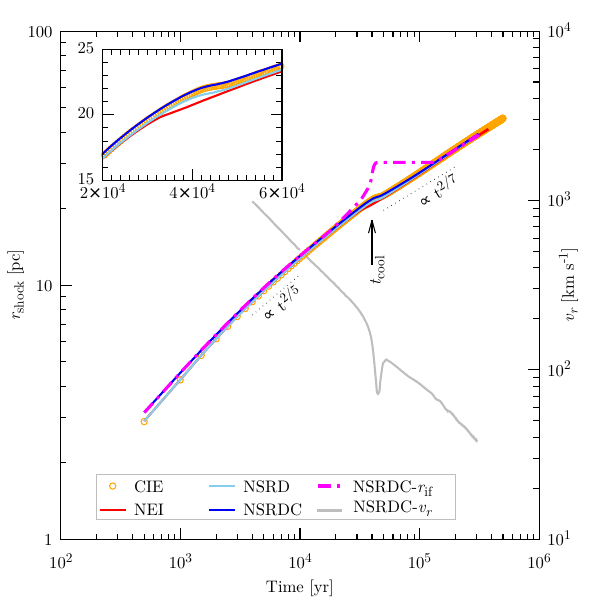}
	\caption{Evolution of shock radius with time for different cases. The cooling time is roughly shown by the vertical arrow. Different line colour shows different cases as described in Table \ref{table:list-of-runs}. The 
	dash-dotted line shows the ionisation front radius only for \textit{NSRDC} case. The inset shows a zoomed in view of the dynamics around the cooling time. The shock velocity for the \nsrdc~case is shown by the grey solid line for a better understanding of the dynamics.}
	\label{fig:Rs-t}
\end{figure}

\section{Characteristics of a supernova}
\label{sec:dynamics}

\subsection{Different phases}
\label{subsec:SN-phases}
The dynamics of a SN remnant is a very well studied problem both theoretically and numerically \citep[][for example]{Cox1972a, Cox1982, Cioffi1988, Kim2015, Steinwandel2020}. The evolution can be divided into the following phases, 
\begin{itemize}
	\item \textit{Free expansion phase:} In this phase the SN ejecta moves freely through the interstellar medium and the radius of the shock front is, $ \rs =$ ejecta velocity $ \times t$. This happens at $t \lesssim 200\: n_0^{-1/3}$ yr, where $n_0$ is the background hydrogen density \citep{Draine2011}. 
	
	\item \textit{Sedov-Taylor phase:} After the end of the free expansion phase, the kinetic energy carried by the SN ejecta is converted into thermal energy by the ISM gas. This energy drives a shock that is self similar in nature and given by the blast wave solution. At this stage, $\rs \propto t^{2/5}$. The observable emission from the earlier part ($\lesssim 5000$ yr) of this phase, however, deviates from a pure CIE plasma and often shows signature of a recombining plasma \citep{Becker1980a, Okon2019}. It is also 
	known that the electron and proton temperatures at this young age of the SN are not equilibrated  \citep{Cox1972, Itoh1978, Cui1992, Ghavamian2007}, a feature that we do not model in our simulations. 
	Our estimation of the emission spectrum during the first few thousand years is therefore likely inaccurate.
	
	\item \textit{Rapid Cooling Phase}: Since the shock slows down with time ($v_s \propto t^{-3/5}$), the shock temperature decreases and at some point it undergoes thermal instability. At this stage, the shock cools rapidly and radiates away most of its thermal energy 
	over a time-scale of $t_{\rm cool} \sim 5\times 10^4\, n_0^{-0.55}$ yr \citep{Cox1972a, Dekel2019}. Although this phase is often considered to be instantaneous, its duration is comparable to the other evolutionary stages \citep{Cox1972a}. Since the physics in this phase has 
	significant consequences on the background material and the shock itself, we put more focus on it. We term the onset of the cooling as $\tcoolon$ ($\approx 20$ kyr for $n_0 = 1\,\pcc$) and the end as $\tcoolend$ ($\approx 50$ kyr for $n_0 = 1\,\pcc$). As a result of 
	this rapid cooling, the shock loses its thermal support and collapses to a very thin shell leaving a hot and low density bubble inside. The shell velocity drops temporarily due to the lack of thermal pressure in the shock. We refer to this structure as `the shell' in our discussion. 
	
	Although cooling is rapid, the creation of the bubble involves two  stages. First, almost $50\%$ of the shocked material (which remains in the outer $\sim 6\%$ of the blast-wave) collapses, thus forming a shell. Second, the remaining material, mostly within 
	$0.8-0.94\: \rs$, cools down at a slightly later time ($\sim 60$ kyr for $n_0 = 1\:\pcc$) because the density of this material is lower. Once this layer has lost its thermal energy, the hot gas pressure of the bubble pushes it towards the already collapsed shell and forms the final shell. 
	The collapse of these two layers creates additional shock-waves, one of which can be seen
	moving inwards through the hot and under-dense bubble (see upper panel of Fig \ref{fig:rho-T-x-vr}).
	
	\item {\textit{Snow-plow phase:}} At $t \gtrsim \tcoolend$, the shell restarts its expansion due to excess pressure from the hot bubble. The expansion of the hot bubble is adiabatic against the background medium, which allows us to estimate the shell radius to be $\rs \propto t^{2/7}$. This is also the period when the shock temperature is $\lesssim \mbox{few} \times 10^5$ K and the cooling time behind the shock is so short that it is practically an isothermal shock . This phase of the SN can be seen in emission lines like \ion{N}{ii}, \ion{O}{iii}, \ion{Si}{ii} \ion{S}{ii} etc \citep{Fesen1980, Ritchey2020a,Ritchey2020b} and is highly susceptible to the effects of non-equilibrium ionisation and photo-ionisation, a primary focus of the current paper. We show the density, temperature and the hydrogen ionisation structure in this phase in Fig \ref{fig:rho-T-x-vr}.
		
	\item {\textit{Momentum conserving phase}:} 
	During the adiabatic expansion, the bubble pressure drops, and at some point falls below the ambient pressure. The shell then enters a momentum driven phase, i.e. $\rs \propto t^{1/4}$. This phase,
	 however, is not often seen in numerical simulations \citep{Cioffi1988} before the shell \textit{fades-away}, i.e. the shell velocity becomes equal to the ambient sound or turbulence speeds, at $t_{\rm fade} \sim 1.9\, \mbox{Myr}\, n_0^{-0.37}$ \cite{Dekel2019}.
\end{itemize}

We run most of our simulations to $300$ kyr and therefore capture the Sedov-Taylor, rapid-cooling, and snow-plow phases. At later times ($t\gtrsim 300$ kyr) the shell velocity drops below $\approx 40\,\kmps$ and it is difficult to distinguish the shell from a turbulent ISM.  We stress that even though the shell is normally considered to be isothermal in the snow-plow phase, this is not necessarily true in our simulations. The increased pressure due to the accumulation of mass in the shell allows it to expand radially inwards, reducing the shell temperature below the background level (see temperature plot at $t = 200$ kyr in Fig \ref{fig:rho-T-x-vr}). 

\begin{figure}
	\centering
	\includegraphics[width=0.5\textwidth, clip=true, trim={1cm 0cm 0.5cm 0.5cm}]{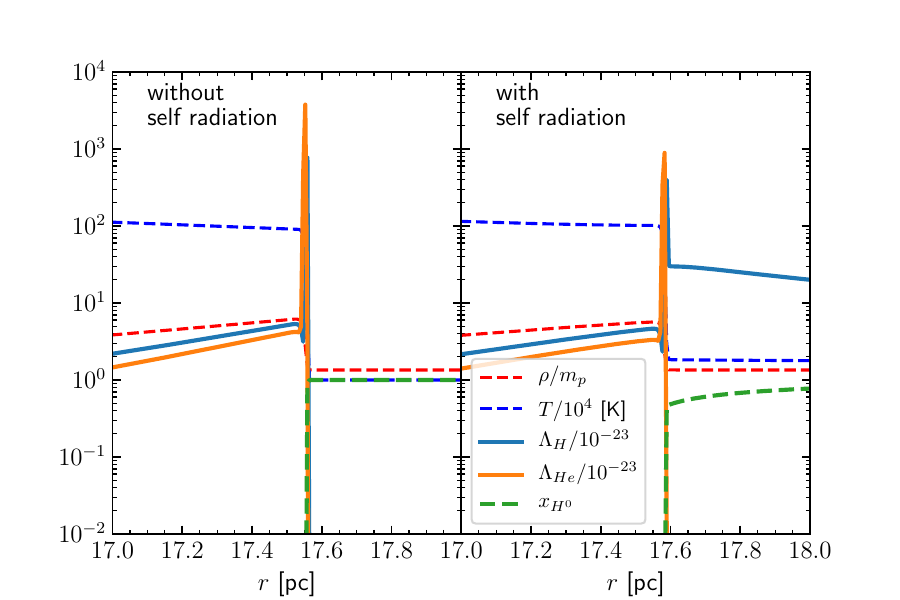}
	\caption{Cooling in the vicinity of shock front at the onset of shell collapse ($t = 23$ kyr) with (right panel) and without (left panel) the self radiation. Different H$^{0}$ and $He^0$ fractions due to different radiation prescription change the elemental cooling rates (shown in units of $10^{-23} \ergps\pcc$).}
	\label{fig:lambda-shock}
\end{figure}

We show the resulting dynamics of our simulations in Fig \ref{fig:Rs-t} for the different physical models. It shows that both the Sedov-Taylor and snow-plow phases are broadly similar for all cases
(table \ref{table:list-of-runs}). 
Discrepancies, however, appear near the cooling time at $\sim 40$ kyr as shown in the inset. This is mainly due to varying H and He cooling (mostly by Ly$\alpha$) at the shock front during the onset at $\tcoolon$. For \cie, the hydrogen fraction, $x_{\rm H^0}$, follows the temperature and hence $x_{\rm H^0} \ll 1$ right at the shock front, whereas, in \nei, hydrogen ionisation time-scale delays ionisation and hence $x_{\rm H^0} \approx 0.9$ at the front.
This increases cooling at the shock front by a factor of few (see the sharp peaks in Figure \ref{fig:lambda-shock}), and hence cooling starts affecting the shock much earlier. Introduction of  self radiation reduces this cooling by photoionising the \HI ahead of the shock and the dynamics falls back towards the \cie~case. With the incorporation of thermal conduction the dynamics becomes almost indistinguishable from the \cie~case. 
Conduction of heat from the shock front further lowers $x_{\rm H^0}$ and, thereby, decreases early cooling by Ly$\alpha$. A similar argument also applies for \HeI cooling.  We therefore conclude from this discussion that \textit{SN dynamics can be well modelled assuming CIE cooling curves without the inclusion of complex physics like NEI, self-radiation and conduction}.

It is clear that the ionisation precursor is of utmost importance in the dynamics of SN around the cooling time. As can be seen in Fig \ref{fig:Rs-t}, the hydrogen ionisation front ($r_{\rm if}$) moves ahead of $\rs$ at $t \approx 20$ kyr ($t_{\rm cool,onset}$). This is when the shell starts cooling rapidly and emits in the UV which can  
ionise the background \HI gas ahead of the shock. The propagation of $r_{\rm if}$ ends at $t \sim 50$ kyr ($\tcoolend$) when the shell has collapsed completely and radiated away most of its energy. This radius can be calculated by equating the total number of ionising photons\footnote{The integration of total ionising photons is done during $22-50$ kyr period. Since part of the photons with energy $>13.6$ eV is also going to get absorbed by \HeI, the available \HI ionising photons = photons($>13.6$eV) $-$ photons($>24.6$eV) + $0.2\times$ photons($>24.6$eV). Here, $0.2$ is the assumed fraction of photons with energy $>24.6$ eV emitted due to direct recombination of He$^+$ to He$^0$ ground state  \citep{Draine2011}. Notice that this is only  $\sim 5$\% of ionising photons if we assume that all the SN energy is emitted in LyC photons. Just to compare, the total amount of radiated energy in the rapid cooling phase is $ \sim 30\%$ of the SN energy (Fig \ref{app-fig:energetics}). Therefore, most of the radiation in this phase is radiated at energies $< 13.6$ eV.}, $N_{>13.6 eV}$ emitted during this period to the total number of hydrogen atoms to be ionised since the source is short lived compared to the \HII recombination time, $t_{\rm rec, H^+}$. The maximum radius of the ionisation front (IF) is, therefore,
\begin{equation}
r_{\rm if, max} = \left( \rs^3 + \frac{3 N_{>13.6 eV} }{4\pi n_0}\right)^{1/3} \,,
\label{eq:rif-max}
\end{equation}
where we have assumed the emission is from a surface with a radius $\rs = 20$ pc (the shock front at $30$ kyr). For our simulation (\nsrdc), the total number of emitted hydrogen ionising photons is $N_{>13.6eV} \approx 2\times 10^{60}$ (see fig \ref{fig:Quv}). This gives $r_{\rm if, max} \approx 29$ pc which is consistent with the maximum IF radius as seen in Fig. \ref{fig:Rs-t}. This estimate for the maximum ionisation front radius is only true for  atoms with ionising potentials above 1 $Ry$. For atoms with lower ionisation potential, like C, Mg, Si, S, Fe the ionisation fronts can be even larger (depending on the spectrum).

After the IF stops propagating, the ionised hydrogen between
$r_{\rm shock}$ and $r_{\rm if, max}$ starts recombining due to the lack of \HI ionising photons. The ionisation front, however, does not change its 
position until it is hit by the shock itself. This happens at $t\sim 130$ kyr when the shell 
reaches $r_{\rm if, max}$ and the velocity of the shell is $\sim 70\, \kmps$. 
By this time, the velocity of the shock-induced ionisation front is smaller than the shock velocity, and a stable radiative-precursor therefore no longer exists. Rather the material that has been ionised at earlier stages lingers ahead of the shock, because it has not yet recombined, and creates an effective ionised precursor to the shock.

\begin{figure}
	\centering
	\includegraphics[width=0.47\textwidth, clip=true, trim={1cm 0cm 1cm 1.5cm}]{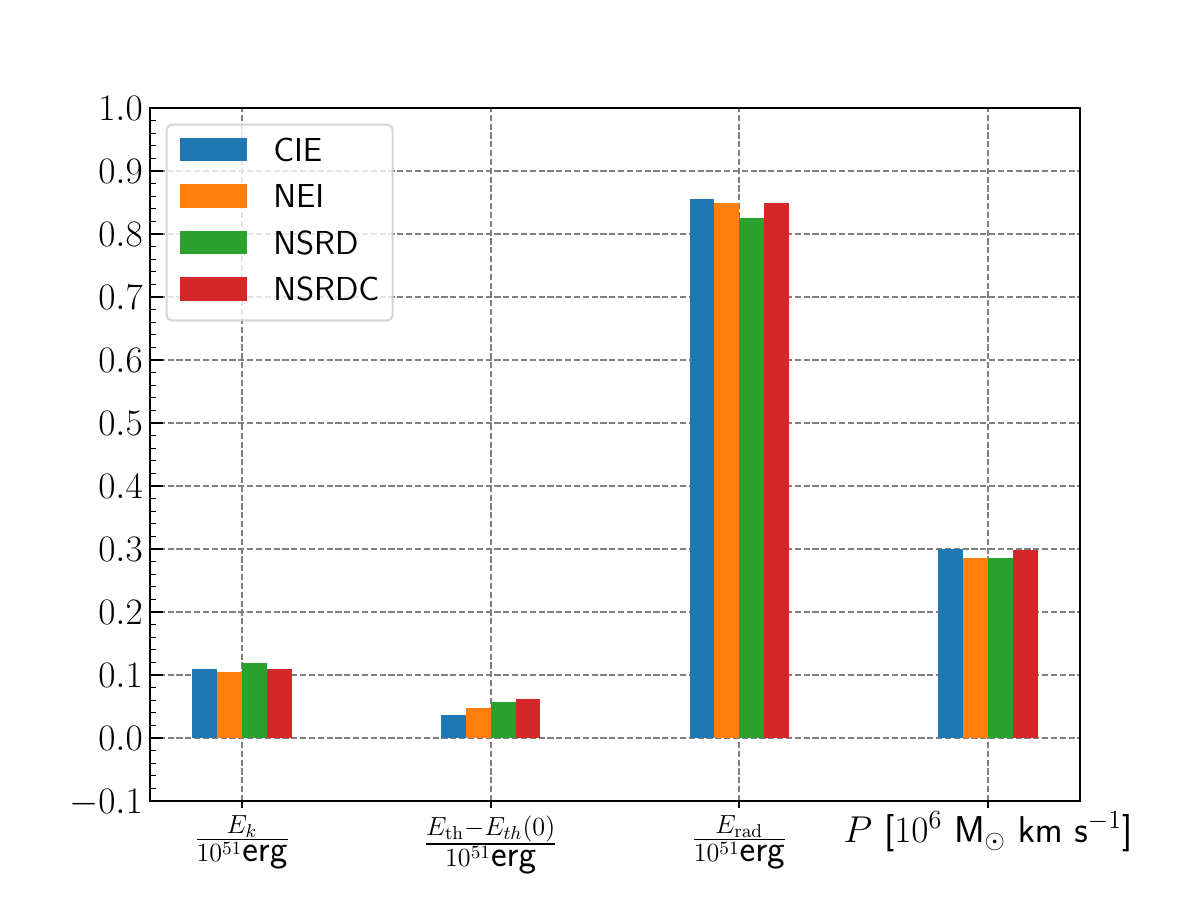}
	\caption{Energetics of a SN in a $n_0 = 1 \pcc$ background medium at the end of the simulation, i.e. $t = 300$ kyr. Simulations containing different physics 
	are shown by different colours. From left to right are the kinetic energy, net gain in thermal energy in the box, radiation loss and total momentum in the simulation box.}
	\label{fig:energetics}
\end{figure}

\begin{figure*}
	\centering
	\includegraphics[width=0.85\textwidth, clip=true, trim={0cm 0.5cm 0cm 1cm}]{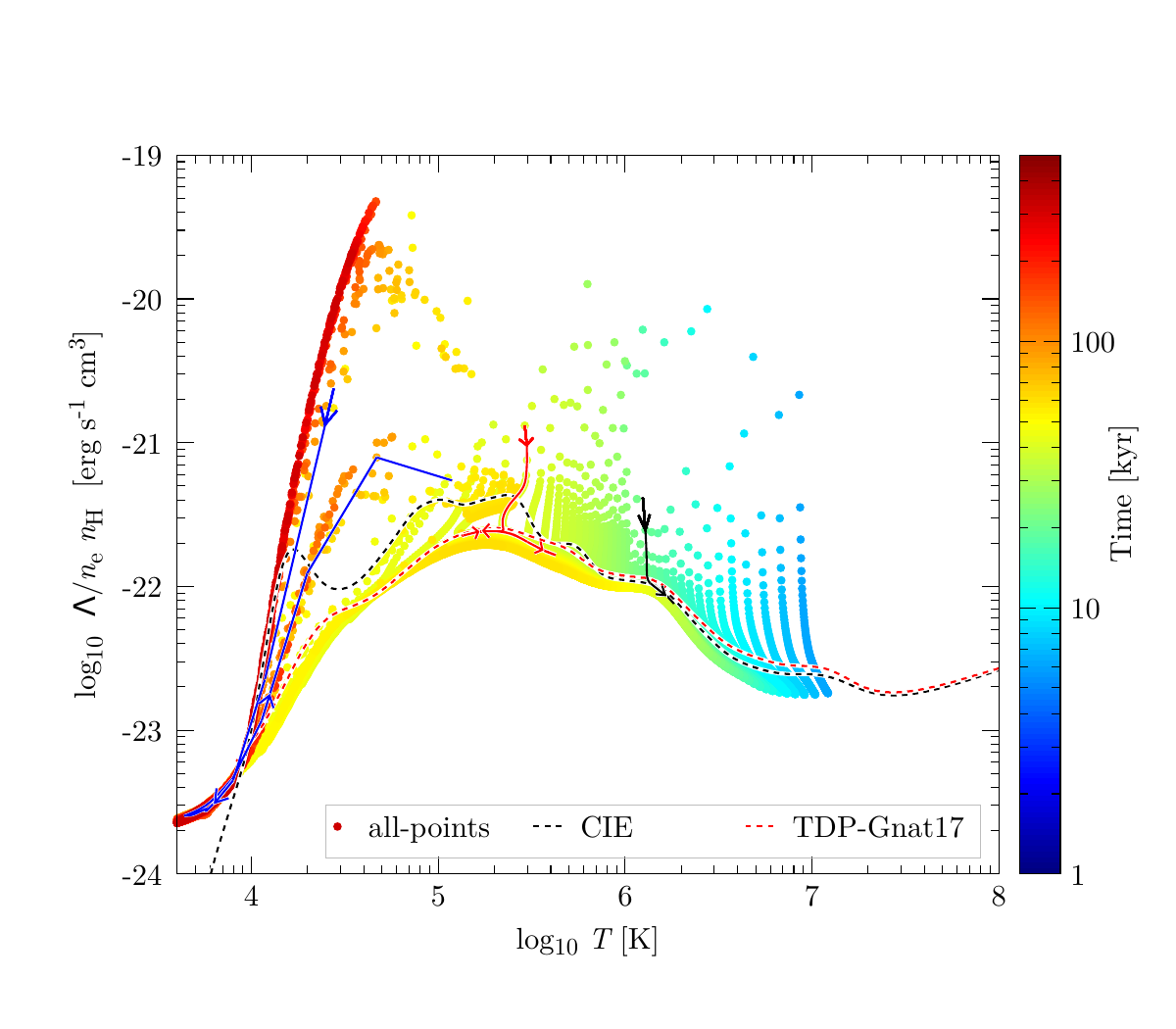}
	\caption{Cooling function in the region between $\rs$ and $r_{\rm bub}$ at all times. Time is shown by the colour of each point. Typical spatial tracks at $20$ kyr (representing Sedov-Taylor phase), $40$ kyr (rapid cooling phase) and $80$ kyr (snow-plow phase) are shown by black, red and blue arrows, respectively. These cooling functions have been compared with more traditional cooling curves. We have not intentionally plotted the points for $t < 5$ kyr since electron temperature may not follow the proton temperature at such early times.}
	\label{fig:cool-curve}
\end{figure*}

\subsection{Energetics}
\label{subsec:energetics-SN}
As has been discussed many times in the literature, the key quantities to know in order to incorporate SN physics in galaxy formation and evolution simulations are the thermal energy, kinetic energy and momentum injection rates for SNe. 
Therefore,  
in Fig \ref{fig:energetics} we plot the total thermal energy and momentum deposited in each case (Table \ref{table:list-of-runs}) at the end of the simulation i.e. $300$ kyr. At this time, the energetics becomes almost constant (see Fig \ref{app-fig:energetics} in the Appendix) and varies by only a few per cent with time. Figure \ref{fig:energetics} shows that 
$80-85\%$ of the SN energy is lost by radiation.
The second column represents the change in the total thermal energy inside the box and and shows that only $\sim 5\%$ of the SN energy is retained as thermal energy of the box. This value goes down further by few \% as time evolves further. The total kinetic energy and momentum retained in the box are  
about $\approx 10^{50}$ erg and $2.8-3.0 \times 10^5 \msun \kmps$, respectively. These numbers are fully consistent with other previous estimates, e.g. \cite{Kim2015}. It is, therefore, clear that \textit{the incorporation of the complex physics like non-equilibrium ionisation or radiative transfer does not change the resulting energetics from the values obtained by simple \cie~simulations by more than a few per cent }.

\subsection{Cooling function}
\label{subsec:cool-func}
As far as the cooling is concerned, it has been suggested that traditional cooling curves obtained from 0-dimensional non-equilibrium isochoric calculations  
can be used as a supplement, instead of the actual ionisation network in more complex hydrodynamic simulations \citep{deAvillez2002, Vasiliev2013a}. In an attempt to understand the working cooling curve behind the shock, we plot the cooling function ($\Lambda/n_e n_H \ergps$ cm$^3$ ) of the shell region (from $r_{\rm bub}$ to $\rs$)\footnote{The bubble radius is defined to be the radius where the density is crosses a certain threshold while going from the centre. This critical value $\ge 0.5 n_0$ during Sedov-Taylor phase and $\ge 5 n_0$ during the snow plow-phase to consider a sufficient region of shocked gas but to avoid including the bubble.} in Fig \ref{fig:cool-curve}. 
The figure also depicts a time evolution of the cooling function, the first $\sim 5$ kyr of which is probably affected by the electron-ion non-equilibrium and, therefore, has not been shown in the plot. We also show two other traditional cooling curves, one for CIE \citep{Gnat2007} and the other is for time dependent isochoric cooling in the presence of photo-ionisation \citep{Gnat2017} for comparison. The figure shows large variations in the cooling functions where it does not follow any known cooling curves. At early times (in the Sedov-Taylor phase, $t \lesssim t_{\rm cool, onset}$, green shades) most of the points, however,  
are concentrated on the CIE cooling curve. This is also demonstrated by the black arrows, 
which represent a spatial track at $20$ kyr, starting from $\rs$ to $r_{\rm bub}$. One can think of this spatial track as the evolution of a single lagrangian cell of gas after it is shocked, and as it flows away from the shock. As the track shows, the cooling for the first few points just behind the shock front is much higher than for CIE and indicates that the plasma is under-ionised. During the rapid cooling phase ($t_{\rm cool, onset}-t_{\rm cool, end}$,orange shades), this curve behaves like a time-dependent photo-ionised plasma (although with a different radiation field). This is
shown by the yellow points and corresponding
spatial track (red arrows). This is due to the highly radiative shell that emits sufficient ionising photons to keep the whole shell close to ionisation equilibrium. The upturn of the arrow towards the end of the spatial track at $80$ kyr represents the bubble-shell interface where temperature rises and the cooling tends to behave like a CIE one.
In the snow-plow phase ($t \gtrsim 60$ kyr, shown using reddish shades), the cooling curve does not follow the traditional curves and rather evolve in a more vertical way. As can be seen in the example track at $t = 80$ kyr (blue arrows), the cooling at the shock front is almost an order of magnitude larger than at CIE. The cooling, however, soon settles down to the CIE values. Below $T \lesssim 10^4$ K, the CIE curve drops sharply due to the absence of any atomic coolant but in the presence of radiation, very tiny amount of ionised H and metals keeps the cooling higher. At even later times
($t\gtrsim 200$ kyr), cooling  
occurs mostly in simple vertical streak at $\sim 2\times 10^4$ K where the cooling just behind the shock can be almost two orders of magnitude higher than the CIE one.
This difference in cooling in the snow-plow phase, however, does not change the dynamics of the shell, since at this point, the shell is mainly driven by the hot pressure of bubble and not the thermal energy of the shell.

\begin{figure}
	\centering
	\includegraphics[width=0.47\textwidth, clip=true, trim={0cm 0cm 0cm 0cm}]{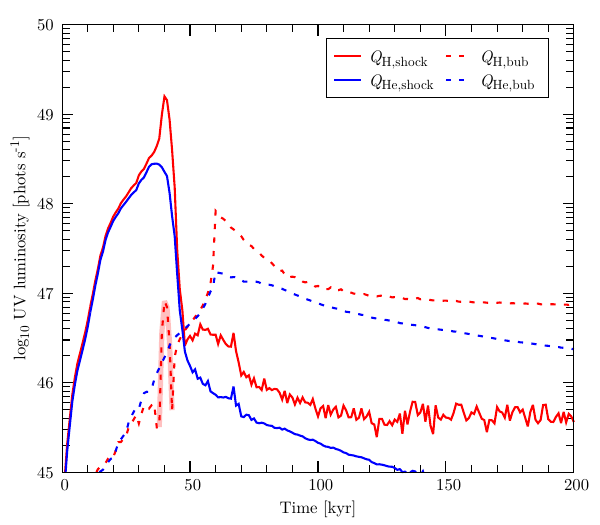}
	\caption{Radial flux of \HI ionising ($Q_{\rm H}$) and \HeI ionising ($Q_{\rm He}$) photons at the shock front ($\rs$) and at the shell-bubble interface ($r_{\rm bub}$). Instances where $Q_{\rm H} < Q_{\rm He}$ mean that there was net radial influx of photons. The shaded part of the curve at $t \sim 40$ kyr is where $Q_{H, {\rm bub}} < 0$  (indicating net influx) and only the absolute value has been plotted to show the values. }
	\label{fig:Quv}
\end{figure}

\subsection{radiative transfer in shell}
\label{subsec:rad-trans-shell}
To understand the nature of precursor ionisation, we show the ionising photon luminosity ($Q = 4\pi r^2 \Phi$ with $\Phi=$ radial photon flux in cm$^{-2}$ s$^{-1}$) from the remnant in Fig \ref{fig:Quv}. We plot the ionising photon luminosity for H ionising ($E > 13.6$) eV and He ionising ($ E > 24.6$ eV) photons just outside the shock front $r = \rs + \Delta r$ (numerically, the next cell) and at the bubble-shell interface $r = r_{\rm bub}$.
As can be seen in the figure, both $Q_{\rm H}$ and $Q_{\rm He}$ rise slowly until $t = \tcoolend$ and then drop sharply after most of the thermal energy in the shock is radiated away. Prior to $\tcoolon$ the photons are mainly He ionising but the photons are mostly H ionising from $\tcoolon$ to $\tcoolend$. This is also the period when the IF detaches from the shock front and pre-ionises the background material. 

The photon luminosities and signs (ingoing or outgoing) at $\rbub$ ($Q_{\rm H, bub}$ and $Q_{\rm He, bub}$) provide an inside view of what is happening at the shock. Let us define a positive luminosity as a net outflux of photons and a negative luminosity as a net influx of photons. The negative luminosity (shaded pink in the background) in $Q_{\rm H,bub}$ between $\tcoolon$ and $\tcoolend$, therefore, means that there is a net influx of photons during this phase due to the extremely bright cooling shell. After $\tcoolend$, the bubble luminosity increases slowly and finally rises above the shock luminosity at $t\approx 45$ kyr signifying negligible emission from the shell. It is clear that the shock luminosity is much smaller than the bubble luminosity at all later times. This means that even if the bubble is emitting some ionising photons, they are
absorbed by the dense shell and only a very small fraction of the ionising photons escape. The escape fraction ($f_{\rm esc}$) of such photons from the shell depends on the time and is $\sim 10\%$ for the LyC photons. The escape of 
He ionising photons is even smaller, of order 
few per cent.

\begin{figure}
	\centering
	\includegraphics[width=0.47\textwidth, clip=true, trim={0cm 3cm 0.5cm 0.5cm}]{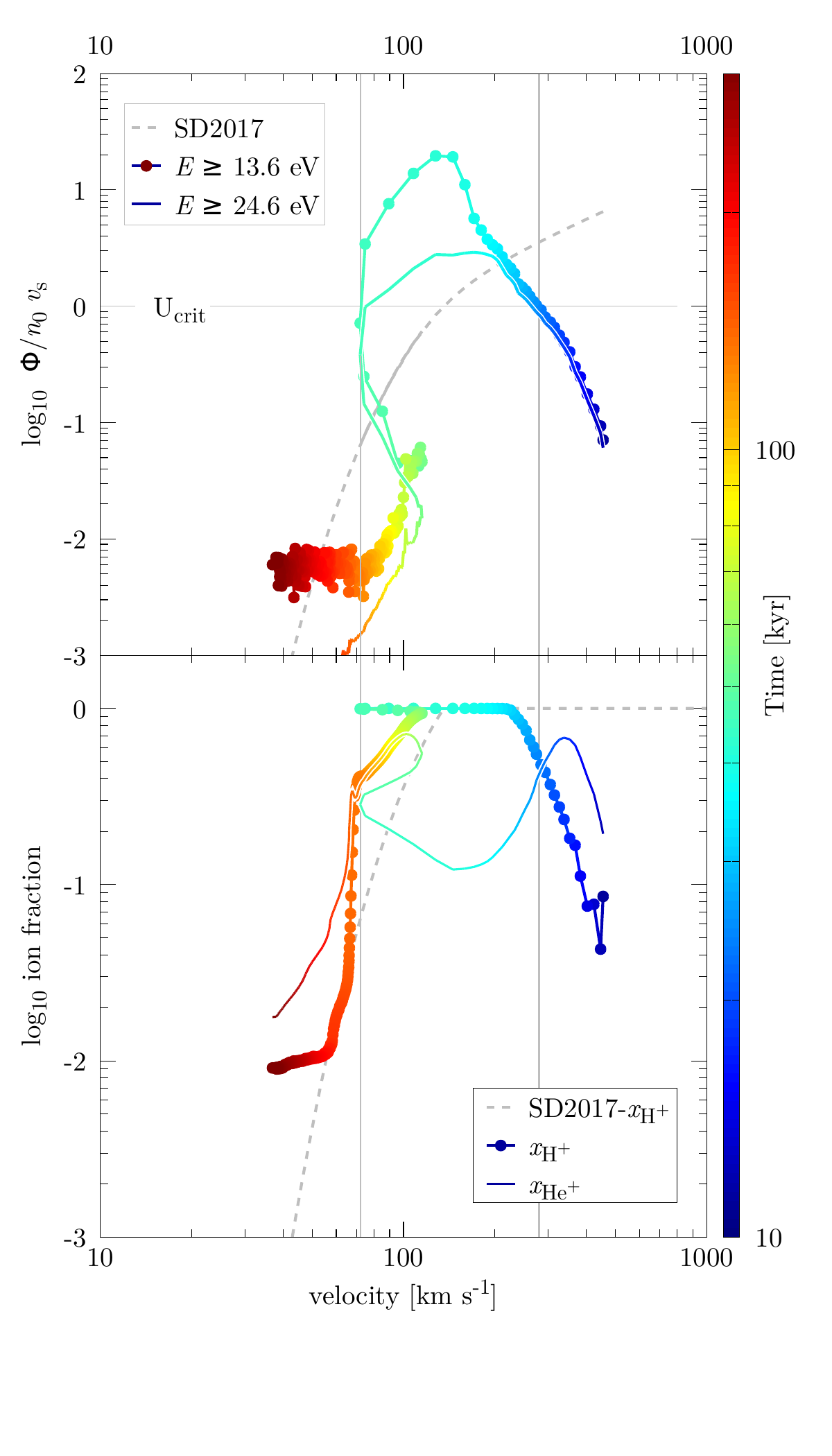}
	\caption{\textit{Upper panel}: Ionisation parameter $U = \Phi/n_0 v_s$ as a function of shock velocity, $v_s$. Colour shows time from the onset of SN. The blue, cyan and reddish part of the curves roughly represent the Sedov-Taylor, rapid cooling phase and snow-plow phase, respectively. The dashed grey line represents the ionisation parameter calculated by \protect\cite{Sutherland2017} (SD2017; their equation 31). The horizontal bar shows $U = U_{\rm crit} = 1$ and the vertical lines show the corresponding velocities. \textit{Bottom panel}: \HII and \HeII ionisation fractions as a function of velocity and time. The dashed grey line shows the calculation from SD2017. The non-monotonous behaviour of the simulated curves with velocity is due to the non-monotonous behaviour of the shock velocity itself during the rapid cooling phase since it stalls temporarily and then restarts again.} 
	\label{fig:vel-Fuv}
\end{figure}

\subsection{Ionisation precursor}
\label{subsec:precursor}
Understanding the ionisation precursor has been at the centre of modelling the emission from shocks. From the pioneering work of \cite{Cox1972} to later works by \cite{Shull1979, Dopita1976, Gnat2009, Sutherland2017} all focused on modelling the ionisation precursor from a plane-parallel shock with a given steady state velocity.
Such models are good for shocks that have already entered the steady state in a background medium that is not affected by anything other than the shock itself. This is certainly not the case for a SN which undergoes different phases (see sec \ref{subsec:SN-phases}) and is  
hardly in steady state.
We therefore present the full evolution of the ionising photon fluxes and H, He ion fractions just in front of the shock (numerically, we choose this to be the cell immediately next  
to the shock front). A useful parameter in this context is the ionisation parameter, $U = \Phi/n_0 v_s$, where $\Phi$ is the photon flux and $v_s$ is the shock velocity. This ratio compares the number of available ionising photons to the incoming flux of neutral gas that needs to be ionised to create a precursor. This means that i) for shocks with $U \ll 1$, the gas ahead of the shock will be only slightly ionised,  and only close to the shock front ii) for shocks with $U \lesssim 1$, the precursor will have noticeable ionised gas but there will still not be any ionisation front, and iii) for $U > 1$, the precursor will be fully ionised and an ionisation front will run ahead of the shock front, forming a radiative-precursor. We term this critical value of the ionisation parameter as $U_{\rm crit}$. 

The ionisation parameter for our \nsrdc~ simulation is plotted in Fig \ref{fig:vel-Fuv} as a function of velocity for a comparison to steady state calculations. Since the shock velocity in the SN remnant decreases over time, it is easier to read the figure from right to left and then to compare it with steady state models (hereafter SSMs). As can be seen in the upper panel of the figure, $U$ initially rises with decreasing velocity due to the decrease in shock temperature and hence increase in overall cooling rate which increases the shell luminosity. In addition, the decreasing temperature of the shock lowers the average energy of the emitted photons (compared to x-ray photons earlier) so that the number of ionising photons increases. It crosses $U = 1$ at a velocity of $\approx 280\: \kmps$, at a time of $\tcoolon$. Note that the main difference between our results and the SSMs at this stage is purely due to the non-steady-state nature of the our shocks. Between $\tcoolon$ and $\tcoolend$ (corresponding a velocity of $280-70\: \kmps$), $U \gg 1$ and the radiative-precursor ionisation front reaches $r_{\rm if, max}$ (see fig \ref{fig:Rs-t}). 

After $\tcoolend$ (at a velocity $\lesssim 70~\kmps$), $U$ drops sharply to $U\approx 10^{-1}$ due to the absence of any photon production from the shell (also seen in fig \ref{fig:Quv}). The shock finally reaches a steady state at $t \gtrsim 70$ kyr  with a velocity $\lesssim 115\: \kmps$. \footnote{Recall that the shock velocity increases from $70\,\kmps$ at $t \approx 50$ kyr to $115\,\kmps$ at $t \approx  60$ kyr due to re-acceleration of the shock after the shell stalls temporarily (Fig. \ref{fig:Rs-t}).} Although $U \ll 1$ at this stage, the gas ahead of the shock is still
slightly ionised, though the ionisation fraction is decreasing with time and velocity (see the lower panel of this figure as well as fig \ref{fig:rho-T-x-vr}). This is because the shock is still inside the ionised sphere ($\rs < r_{\rm if,max}$) created by the rapid cooling phase of the SN, which has not yet recombined. 

The hydrogen ionisation fraction ($x_{H^+}$) finally drops to very low value only at velocities $\lesssim 70\:~\kmps$ when $\rs = r_{\rm if, max}$. From this point onward, the background can be considered  
truly unperturbed. It is an interesting coincidence that this velocity ($70\: \kmps$) is also the limit found by for steady state models below which even a partially ionised precursor cannot be present \citep{Shull1979, Sutherland2017}. 
While our results agree, we differ in physical explanations of this phenomena for a cooling SN shock. In steady state shock models, the precursor is created by photons emitted by the down-streaming material, the characteristics of which depends on the shock velocity. In our evolving-SN shock model, the precursor is put in place during the rapid cooling phase and is independent of the shock velocity afterwards.

The evolution shown in this plot can be divided into four parts. Initially ($v\gtrsim300~\kmps$), the ionisation parameter and ionisation fraction in our evolving shocks are considerably lower than those obtained in the SSMs. This is because our young shocks have only gone through a limited spatial extent, much smaller than the cooling length of the gas. The emitted radiation, which is proportional to the shocked-material depth, is hence much smaller. Later, for $130\lesssim v\lesssim300~\kmps$, both the SSMs and our evolving shocks agree that the ionisation parameter and ionised fraction are significant, although they differ in detail.
At even lower velocities ($70\lesssim v \lesssim 130~\kmps$), the SSMs predict that 
the up-streaming material should only be partially ionised and no fully ionised precursor should form 
\footnote{Note that by `precursor' we refer to a steady ionisation layer that runs ahead of the shock, such as occurs when the ionisation front velocity is larger than the shock velocity. For lower velocities, photoionisation could still occur, but it is limited to the immediate vicinity of the shock front. In this case, the gas may enter the shock ionised, even though a stable precursor does not form. We note that other works do not make this distinction, and refer to both cases by the name  `precursor'.}
\citep{Sutherland2017}. In our models, however, a layer of ionised material still precedes the shock front. This material has been ionised during the rapid cooling phase, and has not yet recombined. As is shown by the grey dashed line in the bottom panel of Fig \ref{fig:vel-Fuv}, that the \HII ionisation fraction produced in this way is larger 
than the expected fraction from a plane-parallel steady-state shock at this velocity range.  Finally, for shock velocities below $70~\kmps$, the ionisation ahead of the shock drops in our evolving models as well and at $60\lesssim v \lesssim 70~\kmps$, the ionisation fraction is consistent with the expected values from the SSMs. Below $v\lesssim 60~\kmps$, the ionisation fraction in our simulation is mostly reminiscent of the initial $10^4$ K ambient gas. \textit{In conclusion},
\textit{the applicability of steady state models to the SN remnants is therefore quite limited.}

\begin{figure*}
	\centering
	\includegraphics[width=0.95\textwidth]{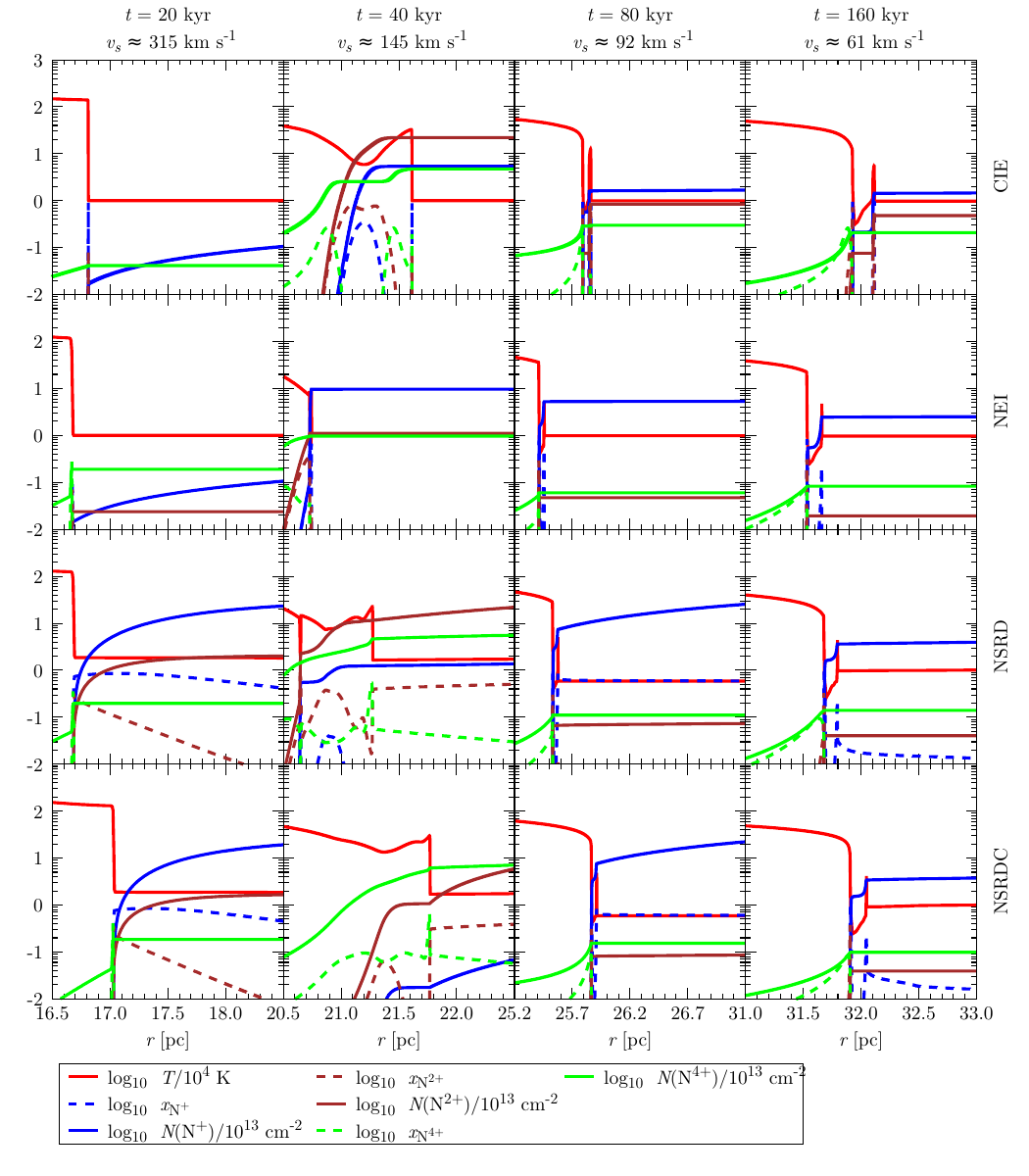}
	\caption{The behaviour of the Nitrogen ions, \nii, \niii and \nv, near the shock front $r = \rs$ for the four model runs with the differing physical processes (rows, legend on the right-hand-side) at four snapshots in the SN evolution (columns). The first, second and third columns represent the Sedov-Taylor, rapid cooling, and snow-plow phases, respectively. The fourth column is for a late time at which the ionisation precursor has vanished.  The solid-red, and the blue, brown and green dashed lines show, respectively, the temperature of the fluid, and the $\xnii$, $\xniii$ and $\xnv$ ion fractions. The corresponding solid lines show the cumulative column densities, $N(<r)$ (in units $10^{13}$~cm$^{-2}$), for the three Nitrogen ions.}
	\label{fig:cumu-col}
\end{figure*}

\section{Observables}
\label{sec:observables}
 We now turn our attention to some of the observable properties of SN remnants which show the importance of these inter-playing physical processes. 

\subsection{Column densities}
Ionic column densities are related to the intensities of emission or absorption lines.
For steady state planar shocks the column densities are functions of velocities. However, as we have discussed earlier, the expanding SN shock is not in a steady state.

Figure \ref{fig:cumu-col} shows the behaviour of example ions near the shock. We focus on \nii, \niii and \nv as our examples. Other ions with similar ionisation energies follow similar trends. For example, \cii, \oii, \sii behave as \nii. The ions like  \ciii and \oiii follow \niii, and ions like \ovi follow the \nv trends. In each panel, we show the temperature, $T$ (red solid line), the \nii ion fraction, $\xnii$ (blue dashed line), and the \nii cumulative column density, $\nnii$ (blue solid line). Similarly we show $\xniii$ (brown dashed), $\nniii$ (brown solid), $\xnv$ (green dashed), and $\nnv$ (green solid). The different rows show simulation results for our four model runs \cie, \nei, \nsrd~and \nsrdc. The columns are snapshots at various phases.
The left most column, at $t \sim 20$ kyr,  
represents the Sedov-Taylor phase when the SN evolution is purely self-similar and the temperatures are high enough for the ions behind the shock front (not at the shock front though) to follow collisional equilibrium. 
Second, at $t \sim 40$ kyr, the shell is in the rapid cooling phase, cooling down to temperatures where non-equilibrium ionisation comes into play. Third, at $t \sim 80$ kyr, the shock is isothermal, but the radiation from the rapid cooling phase is still dominating the precursor ionisation. And fourth, at $t \sim 160$ kyr, the shock is still isothermal but the ionisation precursor has disappeared. As we discussed in section \ref{subsec:SN-phases}, the inclusion of different physical ingredients leads to earlier or later cooling and, therefore, slightly different shock radii as can be seen in the figure. 

\subsubsection{Origin of ions}
\label{subsubsec:origin-of-ions}
Figure \ref{fig:cumu-col} shows that most of the \nii column density in the Sedov-Taylor phase comes from the background, which is either collisionally ionised or slightly photo-ionised due to the radiation from the shock. At later stages 
the region immediately behind the shock-front also contributes to  \nii column. This is because the shock temperature is $\sim$ few $\times 10^4$~K and suitable for \nii production at later times ($t > \tcool$).
Assuming \cie, $\xnii$ follows the temperature.
Therefore, it is large only at the shock front, at the inner boundary of the relaxation layer \footnote{The relaxation layer is defined, in an isothermal shock, to be the layer from the shock front where the gas is just shocked and does not have enough time to cool to the background temperature. The thickness of this layer is roughly $v_s \times $ the cooling time of the shocked gas.} and at the bubble-shell interface where the temperature is suitable to produce \nii. However, the contribution of the bubble-shell interface to the cumulative column density $\nnii$ is negligible.

For \nei, because  the \nii$\rightarrow$ N$^0$ recombination time is longer than the cooling time of the shock, $\xnii$ does not immediately fall to zero inside the shell. Since the density of the shell is very high, even a small \nii fraction  
in the shell can produce a significant $\nnii$. This is the main difference between the \cie~and \nei~column densities.

The introduction of self-radiation in \nsrd~increases $\xnii$ in the shell as well as in the up-streaming material by pre-ionising it. However, the main difference is the presence of \nii in the up-streaming material.
This difference between \nei~and \nsrd~reduces to only a factor of $\sim 2$ at $t \gtrsim 130$ kyr (see section \ref{subsubsec:time-evol-colden}) when the shock reaches the H-ionisation front, $r_{\rm if,max}$ but leaves 
~a trace amount of \nii at larger radii which is still recombining. In addition, since the velocity falls bellow $\sim 70\, \kmps$, it is also not able to create a precursor by itself. Introduction of conduction in \nsrdc~does not change the qualitative picture much from the \nsrd~case, at least for \nii.

Intermediate ions like \niii are mainly generated either at the shock front  or in the precursor region during the Sedov-Taylor phase. For the \cie~ and \nei~ cases, most of the \niii column remains small and originates in the shock-front, whereas in the \nsrd~ and \nsrdc~ cases, the radiation from the shock (which is mostly He ionising before $t< \tcoolon$) 
pre-ionises the upstream material to produce \niii. At later times $t\gtrsim 80$ kyr, the contribution from the shock front can be significant for the \cie~ case. For the cases with non-equilibrium network, the contribution from the shock front is negligible due to the long ionisation time-scale for the \nii$\rightarrow$ \niii transition, and most of the \niii column is formed in the bubble-shell interface. Since the size of this interface and the density and temperature structure at the interface is sensitive to the resolution, this introduces convergence issues for the intermediate ions. We discuss this further in section \ref{subsubsec:convergence}. At still later times ($t \gtrsim 250$ kyr; see figure \ref{fig:origin-of-ions}) the outer region of the bubble cools down to below $10^5$ K and becomes the dominant contributor.

Higher ions like \nv and \ovi originate mainly from the self-similar region in the Sedov-Taylor phase for the \cie~ case, but from the shock front for the non-equilibrium cases. This is due to the long ionisation time-scale for \nv $\rightarrow$ \nvi. A substantially higher column density ($\sim 10^{14} \pcmsq$) of \nv can be observed during the rapid cooling phase when the shell density is higher and the temperature is in the suitable range. The column density decreases in the snow-plow phase for \cie. The main contribution to highly ionised gas then comes from the hot bubble where the temperature is suitable to produce these ions. The introduction of the non-equilibrium effects delays recombination of highly ionised gas (for example, \nvii$\rightarrow$ \nvi, or \oviii$\rightarrow$ \ovii) and hence decreases the ion fractions of the lower ions too. This delayed recombination also reduces the total cooling rate of the hot gas inside the bubble (where $10^5 \lesssim T \lesssim 10^6$ K in figure \ref{fig:cumu-col}). As a result, the bubble temperature is in generally higher in the \nei~ case compared to  CIE. Self-radiation does not affect the higher ions, as the ionisation potentials are much larger than the average photon energy emitted in the snow-plow phase. this is because   the shock temperature in this phase is only a few $\times 10^4$~K.

The introduction of conduction helps to decrease the bubble temperature by transferring energy to the shell, and thereby slightly increases
the recombination rates in the bubble. This causes the \nv fraction and column to increase compared to the no-conduction cases.
This increase in the higher ion column densities, however, is not comparable to the peak achieved during the rapid cooling phase. As is shown in figure \ref{fig:all-colden} which demonstrates that the higher ions like \civ, \nv and \ovi can have column densities $\gtrsim 10^{14}\, \pcmsq$ (for $n_0 = 1\: \pcc$ case) during this phase. \textit{Therefore, observations of these ions in excess of $\sim 10^{14}\: \pcmsq$ may indicate a rapid cooling phase of the SN remnant.}

\begin{figure*}
	\centering
	\includegraphics[width=\textwidth, clip=true, trim={0cm 0cm 0cm 0cm}] {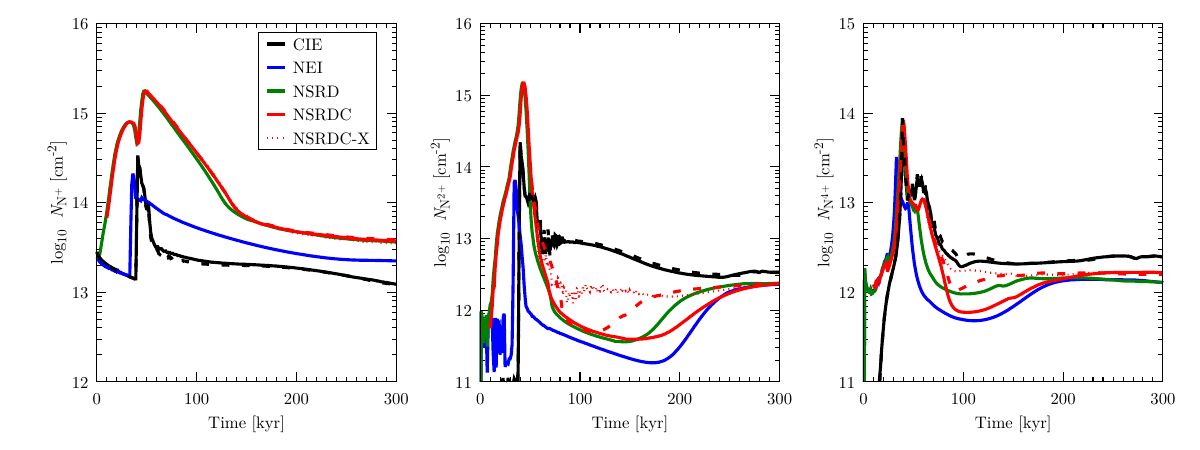}
	\caption{Time evolution of \nii, \niii and \nv column densities from the simulation box. Different cases are
	shown by different colours. The origin of the double-hump structure in $\nnv$ is due to the two stage collapse of the shell during the rapid cooling phase as described in sec \ref{subsec:SN-phases}. The magnitude of the second hump is smaller in non-equilibrium cases due to recombination time of higher ions. The dashed lines show results for higher resolution simulations - $\times 2$ in \nsrdc~and $\times 1.5$ in \cie~case. It shows that \nii is roughly converged but \niii and \nv are not-converged at $\tcool \lesssim t \lesssim 200$ kyr (shock velocity $100 \lesssim v_s \lesssim 50\: \kmps$). This is due to the un-converged temperature structure of the bubble. The dotted line shows a controlled experiment (\nsrdc-X-$0.001$) to achieve converged bubble structure.}
	\label{fig:time-colden}
\end{figure*}
\subsubsection{Time evolution of column densities}
\label{subsubsec:time-evol-colden}
Fig \ref{fig:time-colden} shows the evolution of column densities (integrated from the centre to the edge of the simulation box) as a function of time.  The different curves show the behaviour for our four simulation runs. They all start with an initial \nii column density (left panel) of $\approx 2\times 10^{13} \pcmsq$ corresponding to the initial setup of box size of $50$ pc, density $n_0 = 1 \pcc$ and temperature of $10^4$ K. In the \cie~and \nei~cases, \nii starts to fall as the background material cools down and recombines over time. The column density jumps suddenly at $t\sim \tcool$ due to the rapid cooling phase where the temperature of the whole cooling shell is $\sim$ few $\times 10^4$ K, appropriate for the \nii production. The column density decreases at $t\gtrsim \tcool$ kyr due to a low shell temperature ($\approx 10^4$ K) and the only contribution from the shock comes from the relaxation layer. In \cie~this contribution keeps the total \nii column density slightly higher than the background level. The shell, in the \nei~case, contributes slightly more due to delayed recombination of \nii.

The introduction of radiation has an immense effect on the \nii column at $t \lesssim 130$ kyr ($v_s \gtrsim 70\: \kmps$). During the Sedov-Taylor phase ($t\lesssim \tcoolon$), the He ionising radiation can also ionise N$^0$ to \nii even before the H ionisation front runs ahead of the shock. As explained in the previous section too, most of the \nii column density comes from this precursor region although the contribution from the precursor decreases at $t\gtrsim \tcool$ since \nii recombines with time and finally vanishes at $t\sim 130$ kyr when the shock reaches the H ionisation front. After this point, the main contribution to the \nii column comes from the delayed recombination at the shock front and the trace amount of \nii in the background. 

The slight dip in $\nnii$ at $t\approx \tcool$ happens due to the high number of ionising photons originating from the rapid cooling phase which ionises \nii $\rightarrow$ \niii. This explains why the \niii column (middle panel) peaks rapidly at $t\approx \tcool$. Note that the peak for the cases with self radiation is much higher compared to the normal peak (due to rapid cooling phase) for the cases that do not have radiation (\cie~and \nei). Once the shell cools down completely, the only contribution to the \niii column comes from the forward shock. As can be seen in figure \ref{fig:cumu-col} the shock temperature in the \cie~run is higher than for the non-equilibrium cases. This is because of the under-ionised plasma which increases the cooling efficiency at the shock front as can be understood from Fig \ref{fig:lambda-shock} and \ref{fig:cool-curve}. Therefore, the shock temperatures in the non-equilibrium cases do not reach $\sim 10^5$ K required for \niii production at $t \gtrsim \tcool$. This is why $\nniii$ in the \cie~case is much higher than in the non-equilibrium cases. For the non-equilibrium cases, the main contribution to the \niii column comes from the bubble-shell interface at $\tcool \lesssim t \lesssim 200$ kyr. At $t\gtrsim 200$ kyr, the bubble temperature decreases to $\sim 10^5$ K due to adiabatic expansion and becomes the main source of \niii column. 

It is clear that the peak in column density at $t\sim \tcool$ is a general feature since the shell cools rapidly due to thermal instability and passes through all the temperature zones, thereby producing peaks in all the ion columns. The higher the ionisation potential, the earlier an ion peak appears for (as can also be seen in figure \ref{fig:time-colden}).

Although the behaviour for \nv is similar to \niii, in the sense of a rise to a peak and then a decline, there are some key differences before and after the cooling of the shock. In the Sedov-Taylor phase, the \nv column for the \cie~run is mainly driven by the presence of a small fraction of \nv in the whole self-similar region. On the other hand, in the non-equilibrium cases there are contributions from the self-similar region and also from a thin region at the shock front where the temperature is suitable for \nv production.  This layer is broader in the non-equilibrium cases due to the ionisation time-scale for \niv $\rightarrow$ \nv. This leads to a larger \nv column even in the Sedov-Taylor phase for the non-equilibrium cases.
After $t \gtrsim \tcool$, the temperature in the bubble becomes suitable for \nv production and, therefore, becomes the main source of \nv. The non-equilibrium cases, on the other hand, contain a low ionisation fraction, $\xnv$, due to the delayed recombination time of higher ions as explained in section \ref{subsubsec:origin-of-ions}. Conduction at the shell-bubble interface helps transfer extra heat from the bubble to the shell and mass from the shell to the bubble, thereby increasing the recombination rate of highly ionised nitrogen. The actual amount of \nv contribution from the bubble, however, depends on the exact temperature and density values of the bubble at any time, since the the recombination rate is a steep function of temperature in this regime. 

We conclude that line emissions from high ions such as \nv in the Sedov-Taylor phase are expected to trace under-ionised gas (due to the ionisation time of \niv$\rightarrow$\nv), and indicative of non-equilibrium effects.
Similarly observing over-ionised line emission after $t \sim \tcool$ would be a definitive probe of non-equilibrium physics inside the bubble (due to the long recombination time of \nvii$ \rightarrow$ \nvi $\rightarrow$ \nv).

\subsubsection{numerical convergence}
\label{subsubsec:convergence}
We check the convergence of the ion column densities for the \cie~ and \nsrdc~ cases only.
Since the \nsrdc~model has the largest number of physical ingredients included,  
we plan to compare its column densities with both 
steady state models and the \cie~case. The dashed lines in Fig \ref{fig:time-colden} show higher resolution simulations with resolutions $\times 2$  
for the \nsrdc~case and $\times 1.5$  
for the \cie~ case. Although \nii seems 
converged, \niii and \nv are  
clearly not converged between $\tcool \lesssim t \lesssim 200$ kyr. The main reason is the slightly different temperature values in the bubble at different resolutions and the unresolved bubble-shell interface. Since both \niii and \nv  
are produced either in the interface or the bubble (and because $\xnv$ is highly dependent on the bubble temperature), these columns remain sensitive to the resolution.

\begin{figure}
	\centering
	\includegraphics[width=0.45\textwidth, clip=true, trim={0.5cm  1.6cm 0.5cm 0.5cm}]{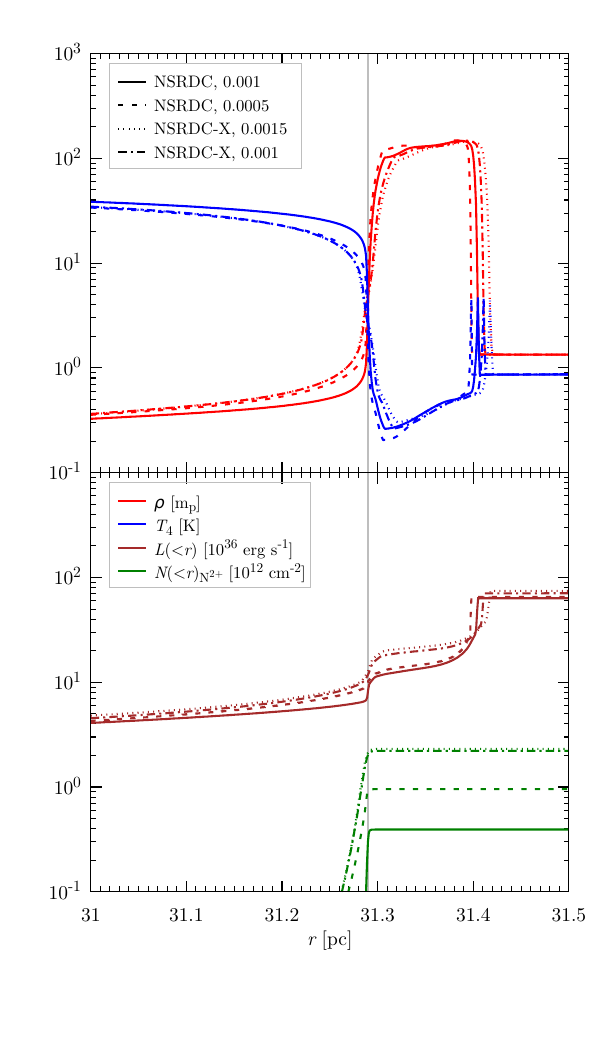}
	\caption{Checking numerical convergence with controlled experiment on \nsrdc. The figure shows density (red), temperature (blue), cumulative bolometric luminosity (brown) and cumulative \niii column density at $t = 150$ kyr. The solid lines represent \nsrdc~($\Delta x = 0.001$ pc), the dashed lines show higher resolution version of \nsrdc, the dotted lines represents results from \nsrdc-X, $\Delta x = 0.0015$ pc and the dot-dashed line represent results from \nsrdc-X with $\Delta x = 0.001$ pc. We have adjusted the structures in $r$-direction slightly ($\sim 0.02$ pc) so that the interfaces of all the runs match at a given location for better visibility. The fact that the dotted and dot-dashed lines are hardly distinguishable in the bubble region indicates excellent convergence in the \nsrdc-X case.}
	\label{fig:convergence}
\end{figure}

Although the non-convergence of the bubble and the bubble-shell interface appear unrelated,
the bubble is in fact affected by the property of the interface. To clarify this statement, let us examine the density and temperature structures of the region near the interface. We do this for the \nsrdc~case at $t = 150$ kyr, where both $\nniii$ and $\nnv$ seem to be un-converged. This is shown in Fig \ref{fig:convergence} where the density is shown in red, temperature in blue, bolometric luminosity, $\mathcal{L}(<r)$ in brown and the \niii column  in green. The solid lines show the values for $\Delta x = 10^{-3}$ pc and the dashed lines show the results for $\Delta x = 5\times 10^{-4}$ pc resolution. Clearly, the density and temperature of the bubble are not converged. This  
is very apparent close to the interface which is marked by the gray vertical line. 
The main reason for this non-convergence is the 
un-converged cooling at the interface. Radiative cooling at the interface  
causes the local plasma to lose its thermal support 
thereby accumulating on to
the shell. Therefore, higher cooling at the interface leads to higher mass accumulation rate from the bubble to the shell which in turn means that the bubble becomes less dense. 
The lower panel of Fig \ref{fig:convergence} shows that the jump in the bolometric luminosity, $\mathcal{L}(<r)$ is smaller at higher resolution, meaning that the interface undergoes less cooling at higher resolution owing to a thinner interface region. This also affects the total energy radiated by the remnant as can be seen in Fig \ref{app-fig:energetics}.

Ideally, for an infinite resolution we would be able to resolve the thickness of the interface (Field length is $\sim 3\times 10^{-6}$ ~pc), which, compared with our resolution element, is very small. This implies negligible
cooling loss at the interface and negligible  
mass accumulation
from the bubble to the shell. This would thus allow the bubble to conserve its mass. Since the simulations performed here are computationally expensive, we do not perform even higher resolution simulations. Instead, we perform normal resolution \nsrdc~simulations, but artificially turn off radiative cooling at the interface\footnote{We define the interface ($r_{\rm interface}$) to be the first instance encountered where the density $> 5 n_0$ while going out radially from the centre. The zone, where the cooling is prevented,
is defined to be the region that lies within $r_{\rm interface}\pm 4 \Delta x$. Here $\Delta x$ is the resolution of that particular simulation.} while allowing the ionisation network to operate normally. The prescription is only used at $t \ge 60$ kyr (snow-plow phase) when the bubble-shell interface becomes apparent. We call this series of simulations as \nsrdc-X (\nsrdc-experimental).

The results for the \nsrdc-X simulations are also shown in Fig \ref{fig:convergence} by the dotted ($\Delta x = 0.0015$ pc) and dot-dashed ($\Delta x = 0.001$ pc) lines. The fact that these two lines are hardly distinguishable in the bubble region shows the success of this experiment. The result is also intuitive. In the absence of any mass accumulation from the bubble to the shell, the bubble now contains the maximum amount of mass possible after the rapid cooling phase. Since the bubble pressure is only a function of time, the temperature of the bubble at any given time is now the lowest possible. The \niii column density in the bottom panel (green dotted and dot dashed lines) also show the convergence for the experimental runs.

Our zero cooling prescription at the interface leads to an artificial density and temperature profile that may differ from the hypothetical infinite resolution simulation. Therefore, any contribution towards the ion column density from the interface region is poorly represented.
We present the column densities of the \nsrdc-X-$0.001$ pc in Fig \ref{fig:time-colden} after subtracting the contribution from the interface. Although we do not show the \nsrdc-X-$0.0015$ pc column densities to avoid overcrowding, we verified that the they are converged (as implied by 
Fig \ref{fig:convergence} too). 

In the following comparison of our data to the steady state shocks, we quote the \nsrdc-X-$0.001$ pc results.

\begin{figure*}
	\centering
	\includegraphics[width=0.9\textwidth, clip=true, trim={0.5cm 0cm 0.5cm 0 cm}]{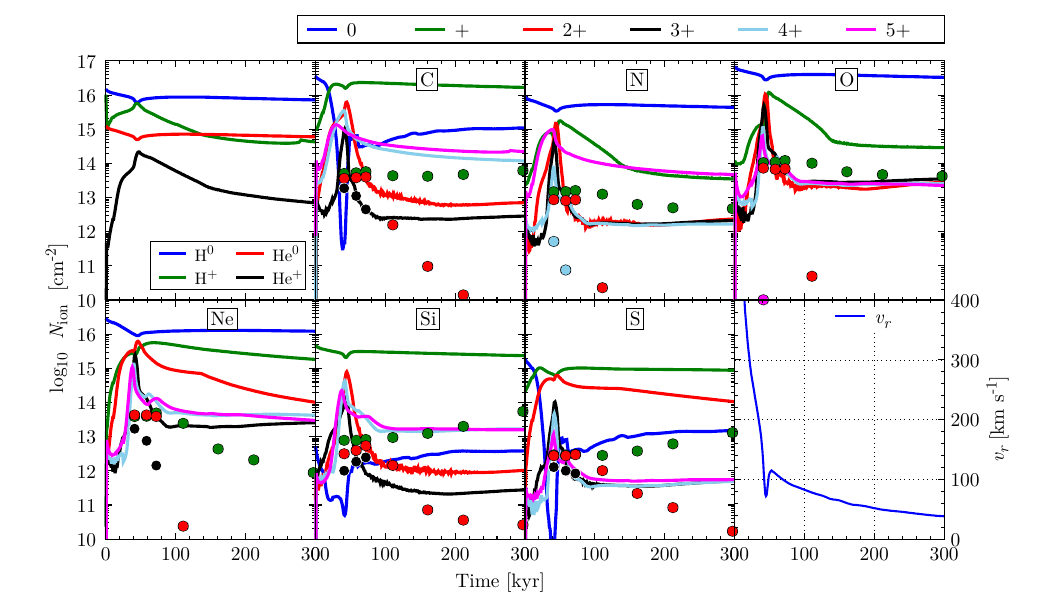}
	\caption{Time evolution of different ion column densities (integrated till the shock front) in the case of \nsrdc~($n_H = 1.0 \pcc$). Different panels show different species, whereas different lines in each panel represent ions of that species. The top-left panel shows H and He column densities after dividing by $10^4$ to bring them to the same scale as others. The small break at $t=60$ kyr is due to the subtraction of contribution from bubble-shell interface. The bottom-right panel plots the shock velocity with time for reference. Machine readable table containing the simulated column density can be found in the online version.}
	\label{fig:all-colden}
\end{figure*}

\subsubsection{Comparison with steady state models}
\label{seusubsec:comparison}
The evolution of the important ion stages for each species (except Mg) for the \nsrdc-X-p001 case  is shown in figure \ref{fig:all-colden}. Each panel is for a different element, and the curves show the time evolution for each ion of individual elements.
The evolution is shown as a function of time rather than velocity since a large part of the evolution is not in steady state. In fact, near $t \sim \tcool$ the velocity is double valued since the shell stops temporarily due to the lack of thermal pressure in the shock but restarts it journey once the hot bubble starts expanding. However, conversion to instantaneous velocity can be easily done using the time-velocity plot shown in the bottom-right panel. Figure \ref{fig:all-colden} shows that the metal ions are dominated by the neutral or first ionisation stages present either in the thin shell or in the background. The flat behaviour for the lower ions like, C$^0$, C$^+$, Si$^0$, Si$^+$, S$^0$ and S$^+$ reflects the fact that they are produced mostly in the background (region outside the H ionisation front). The sudden fall of atomic column density of 
C$^0$, Si$^0$ and S$^0$ is due to the photo-ionisation caused during the rapid cooling phase ($\tcoolon \lesssim t \lesssim \tcoolend$). \footnote{Note that the ionisation front for C and S extends beyond our computational box of $50$ pc. In addition, elements like Mg, Si, Fe are already singly ionised by the assumption of our initial condition at $T = 10^4$ K. Therefore, the column density of \cii, Mg$^+$, S$^+$ and Fe$^+$ are only a lower limit as it depends on the box size.}. As explained earlier, this is also the reason why the column densities of the next ionisation level for these elements are higher at this time. After the rapid cooling phase is over, the singly ionised atoms recombine according to their corresponding recombination rate coefficients and the local electron density. For example, the recombination time-scale for \cii $\rightarrow$ C$^0$ is $t_{\rm rec, C^+} \sim 1/\left( n_e\:\alpha_{C^+}\right) \approx 40/n_e $ kyr, assuming $\alpha_{C^+} \approx 8 \times 10^{-13} \psec$ cm$^3$ at $T = 10^4$K.  This means that \cii can recombine at a time-scale of $\sim 40$ kyr within the hydrogen ionisation front where $n_e \sim 1$ which is also seen in the increasing column density of C$^0$ until $\sim 130$ kyr when the shock reaches the H-ionisation front. At larger radii, carbon mostly remains singly ionised since $t_{\rm rec, C^+} \gtrsim$ Myr owing to the very low value of electron density ($n_e \lesssim 10^{-2}$, given H is mostly neutral at this region).  

We also compare our results with steady state shock calculations. We use the steady state, plane-parallel shock models presented in \cite{Gnat2009} that contain self-consistent radiation field but re-run their models for lower velocities ($\sim 40-150\: \kmps$) where the SN remnant spends most of its lifetime. The resulting comparison is shown in Figure \ref{fig:all-colden}. The coloured points show the column densities estimated from the models of \cite{Gnat2009} at `given shock velocities' but converted to `given times' by using the time-velocity curve of the simulated SN shock. Clearly, the shock velocities used for steady state shock calculations are only achieved for a very short duration of time in a realistic SN shock and, therefore, it does not get enough time to set up a structure similar to a steady state structure which makes the comparison a bit unfair. However, since many of the SN studies consider such steady state models (SSMs) to infer either the velocity or the metallicity of the ISM \citep[for example][]{Dopita1980} it is worthwhile to compare these two cases. 

Fig \ref{fig:all-colden} shows that the SSM  column density estimates for the lower ions like \nii, \oii, \oiii etc can be off by almost an order of magnitude at $t\lesssim 130$ kyr ($v_s \gtrsim 70\: \kmps$) in the presence of a recombining precursor. At later times, this discrepancy comes down to only a factor of $\sim 5$. The discrepancy is even higher for intermediate and higher ions like \niii, \nv, \ovi where the simulated results are almost 2 orders magnitude higher than the expected values from SSMs. Such a severe underestimation of these ions in SSMs is simply due to the fact that the SSMs do not have a hot/warm bubble where most of these ions are produced. A slightly better agreement for the lower ions is expected since the SSMs are roughly able to capture the evolution of density,  temperature and ion fractions just behind the shock where a good fraction of these ions are produced. 
We conclude from this part that \textit{the column densities obtained for a SN remnant using steady state models do not represent the observable column densities}.

\begin{figure*}
	\centering
	\includegraphics[width=0.95\textwidth, clip=true, trim={0cm 0cm 0cm 0cm}]{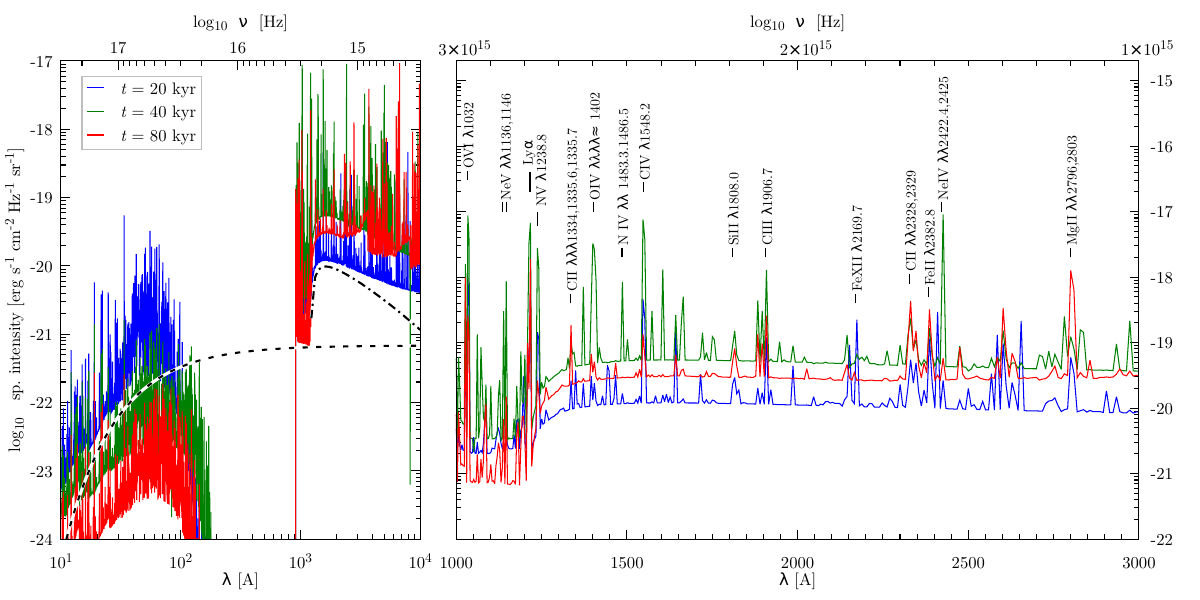}
	\caption{Emission spectra at $t = 20$ kyr (blue), $t = 40$ kyr (green) and $t = 80$ kyr (red) for the \nsrdc~run at zero impact parameter. The spectra are obtained by performing full radiative transfer till the edge of the simulation box, i.e. $50$ pc. The left panel shows the full spectra from $10-10,000$ \AA, whereas, the right panel shows a zoomed in version of it to indicate a few possible emission lines. The dashed black line in the left panel shows arbitrarily scaled Bremsstrahlung emission from $T = 2\times 10^6$ K plasma \citep{Draine2011} and the dot-dashed black line shows arbitrarily scaled two-photon emission spectra \citep{Nussbaumer1984} for comparison.}
	\label{fig:SN-spectra}
\end{figure*}
\subsection{Emission spectra}
In Fig \ref{fig:SN-spectra} we present emission spectra at different times for the \nsrdc~run. We computed the spectra with high resolution (HR) radiative transfer using the methodology described in paper I. We include a total of $3457$ frequency bins ranging from $10^{13}$ Hz ($30 \mu$m) to $10^{18}$ Hz ($3$ \AA) with a frequency resolution $\mathcal{R} = \Delta\nu/\nu \approx 298$. Opacities and emissivities are obtained from \cloudy-17.01 at each individual radius given the local density, temperature and non-equilibrium ion fractions of the cell. Although we solve the on-the-fly radiative transfer using only $16$ rays ($-1\le \mu \le +1$), we employ a total of $128$ rays for our HR computation. The large number of rays allows us to probe the spectra at different impact parameters across the SN remnant. The impact parameters ($b$) are simply converted from the $\mu$ values at that radius since $b = r \sin\theta = r\:\sqrt{1-\mu^2}$.
We calculate the spectra at the edge of the simulation box assuming that any additional emission or absorption by the foreground medium can be calculated easily.
	
We show the resulting spectra for $b=0$ in Fig \ref{fig:SN-spectra} for three different times representing different phases of the SN remnant. Typical emission from a plasma in our case consists of the free-free emission (Bremsstrahlung), the free-bound emission (continuum emissions from recombination of free electrons onto ions, and the bound-bound transition (mostly line emissions, but can also be continuum emission in special cases). The absorption is mostly from the Ly-continuum and different metal-lines. 
As can be seen in the left panel of the figure, the x-ray emission at $\lambda \lesssim 100$ \AA is due to the Bremsstrahlung emission from hot plasma at the shock front and almost vanishes at later times when the shell is much cooler. The dip in the spectra around $\lambda \sim 100-912$ \AA is due to the Ly-continuum absorption by the neutral hydrogen present in the region between the shock front to the edge of the simulation box.
The bright emission lines immediately after $\lambda = 912$ ~\AA are the Ly-$\delta$, Ly-$\gamma$, Ly-$\beta$ lines from recombining hydrogen plasma. Another bright peak clearly seen at $1216$ \AA is the Ly-$\alpha$ emission line. Note that the actual line brightness and the line shape of the Ly-$\alpha$ line may be different than obtained here since this is a highly resonant line and it diffuses both in space and frequency which are not modelled in our radiative transfer algorithm. The continuum emission right after $\lambda \ge 1216$ \AA is the two photon emission from hydrogen $2s\rightarrow 1s$ level (since a single photon emission is forbidden between these two levels).

We show a zoomed in view of the spectra in the right panel of the Fig \ref{fig:SN-spectra} in the UV frequency range that can be probed by HST COS or other spectrographs. We find that emissions from higher ions like, \ion{O}{iv}, \ion{O}{vi}, \ion{N}{iv}, \ion{N}{v} are only present at earlier time ($\lesssim 40$ kyr) when the shock temperature is higher. The lines are brighter at $t \approx  40$ kyr since the temperature of the shell is $\sim 10^5$ K, suitable to produce these ions.  At later times ($t\gtrsim 80$ kyr), emission from the lower ions like, \ion{C}{ii}, \ion{Mg}{ii} etc prevail. As can be seen in the left panel that the SN remnant becomes much brighter in optical bands ($\sim 4000-7000$ \AA) at later times as the shock slows down.
\section{Discussion} 
\label{sec:discussion}
In this paper we have studied the effects of non-equilibrium ionisation and associated cooling, radiative transfer, and thermal conduction on the structures and chemical states of expanding supernova blast wave shells.  There are several physical, numerical as well as methodological aspects that we have not included so far. We discuss them here.

\subsection{3D vs 1D}
\label{subsec:3d-1d}
One important difference between our simulations and real SN remnants is our assumption of spherical symmetry. It is well known that a  thin shell driven by thermal pressure of the interior gas is subject to shell instabilities such as the Vishniac instability \citep{Vishniac1983} when the assumption of spherical symmetry is dropped and the shock is radiative enough so that the effective adiabatic index is $\lesssim 1.1$ .
This instability arises even if the shock is moving into an initially uniform background. It occurs because the hot gas pressure acts normally to the surface, whereas ram pressure on the shell presses radially inwards. This causes the shell to have a net force along the non-radial directions and any small deformation of the spherical shell results in filamentary structures and finally fragmentation \citep{Krause2013, Steinberg2018}. Therefore, a full 3D or at least a 2D simulation may produce significant clumping of the shell, thereby breaking the spherical symmetry. Such clumping may increase the dust opacity in the shell by reducing the dust destruction. \cite{Steinberg2018} have shown that the average shock temperature where most of the cooling occurs may also change due to such clumping. The average shock radius, however, does not seem to vary much compared to a 1D calculation \citep{Krause2013, Yadav2017}. 
However, analysis of laboratory shocks \citep{Grun1991, Laming2002, Laming2003} suggests that atomic shocks in cosmic abundance plasma are unlikely to be sufficiently radiative.

In addition to the 3D instabilities, pure 1D instabilities, such as oscillatory instability also may arise in the thin shell \citep{Chevalier1982}. It is expected that the radiative shocks, following a cooling function $\Lambda(T) \propto T^\alpha$, are generally stable against the 1D oscillatory instability if $\alpha > \alpha_{\rm cr} = 3$. Although later works show that this critical exponent, $\alpha_{\rm cr}$, is dependent on the Mach number of the shock and whether the shock needs to cool down only to the level of the pre-shocked material or not \citep{Laming2004, Pittard2005}. These authors show that for $\alpha = 3$, the radiative shock becomes unstable only at Mach $\gtrsim 10$. With lower Mach numbers, the $\alpha_{\rm cr}$ decreases. \cite{Pittard2005} show that for Mach $= 5$, $\alpha_{\rm cr} = -0.4$ at least for the first overtone (which contains most of the power) and the system becomes stable against the oscillatory instability. Given that a roughly steady radiative shock only appears in our simulations at $t \gtrsim 60$ kyr i.e., $v_s \lesssim 110~ \kmps$ (Mach $\lesssim 10$), Fig \ref{fig:cool-curve} suggests that $\alpha \gtrsim 2$ at these velocity range. Therefore, we speculate that in the case presented here, the radiative shock is mostly stable against the 1D oscillatory instabilities. However, a definite answer to the question may be difficult to provide in our current simulation set-up since we do not impose any perturbations on the expanding shell. Besides, the cooling layer behind the shock may not be well resolved due to our numerical resolution.

Throughout our simulations we have assumed that the background material is uniform in density and pressure owing to the uniform temperature. In reality, the ISM density if often non-uniform and much of the pressure comes from the non-thermal components, like, turbulence. For example, the the typical turbulence speed in the Milky-Way ISM is about $\sim 15-20 \,\kmps$ \citep{Krumholz2016} compared to the adiabatic sound speed of $\approx 15\, \kmps$ (assuming $T = 10^4$ K). Clearly, the turbulent structure in the ISM is non-negligible.  This can further amplify fragmentation.

\subsection{Mixing Layer prescriptions}
As we explained in section \ref{subsubsec:convergence}, the formation of ions such as \niii and \nv are heavily dependent on the bubble properties which depends on the exact amount of cooling and accumulation of gas at the bubble-shell interface. The interface itself is a significant source of ions like \niii, \oiii etc. The 3D instabilities in the  
interface region, therefore, can substantially change these ion columns. Addressing this issue is out of the scope of the current paper. One thing that we could do in the current simulation is to use a mixing layer prescriptions \citep[for example,][]{Duffell2016} as used by \cite{ElBadry2019}. We, however, must note that the validity of these prescriptions are verifies 
only in controlled cases, and  
only in terms of the average density and temperature profiles. Given that the total cooling and ion columns strongly depend on the local density, temperature and the evolution history, a traditional mixing layer prescription seems insufficient and may even lead to completely wrong column densities. We therefore do not use (or plan to use) a mixing layer prescriptions in our work.

\subsection{Magnetic field}
\label{subsec:mag-field}
In our study of the effects of non-equilibrium ionisation/recombination with the self-radiation, we have not included any magnetic field (MF; $B$). \cite{Slavin1992} found that for a SN remnant expanding into a medium with $n_H = 0.2 \pcc$ a magnetic field  suppresses the density jump behind the shock and depends on the exact strength of the field. The shock radius seems to grow faster at $t \gtrsim 1$ Myr. In a recent simulation,  \cite{Evirgen2019} showed that for a SN into a $n_H = 1 \pcc$ medium, the shock radius depends on the direction of the magnetic field. For a shock moving perpendicular to the field, the difference can be only $\sim 5\%$ at $t \approx 400$ kyr (for $B = 5 \mu$G), whereas the shock radius does not change much if the shock moves parallel to the field direction. They also showed that the residual net energy (both kinetic and thermal) only increases by a factor of $20\%$ compared to no MF cases after the end of $500$ kyr. We, therefore, do not expect any dramatic changes in the dynamics of the SN remnant simulated here.

The column densities and emission spectra are, however, expected to change due to the presence of the MF \citep{Gnat2009, Patruk2016, Bach2019}. 
The density and pressure profiles behind the shock can be altered during the rapid cooling phase (their figure 4). This may cause a lower radiation in high field strengths. Since the precursor region in our simulation depends on the radiation from the rapid cooling phase, we expect this region to be smaller for increasing MF. \cite{Patruk2016} also showed that the suppression in density jump behind the shock leads to an increased shock temperature in general. Since both the cooling function behind the shock and the recombination rates are dependent on the actual density, non-equilibrium ion fractions and local radiation field in a non-trivial manner (Fig \ref{fig:cool-curve}), it is hard to speculate the exact outcome of the MF on the column densities and emerging spectra. We therefore keep this task for future work.

In addition to the direct effects of the magnetic field on the shock, a shock running through a magnetised ambient medium may also accelerate particles and plasma waves which may radiate away thermal energy of the shock via non-thermal channels \citep{Ediston1984, Quest1988, Laming2001, Treumann2009}. Therefore, this process can make the shock prone to instabilities (discussed in previous sections) by increasing cooling. It is also possible that the particle acceleration and plasma waves can amplify the magnetic field upstream, thereby increasing the stability of the shock against the above mentioned instabilities. One needs to perform full MHD simulations (preferably using `particle-in-cell method') along with non-equilibrium ionisation and radiative transfer to properly answer these question.

\subsection{Dust grain size}
In our computations we make the simplifying approximation that the dust extinction cross section is proportional to the area of a typical dust grain of initial size $a = 0.1\: \mu$m.
We do not consider a grain size distribution. The effect of dust is, however, expected to change our results only for the high density runs as we discussed in paper I. The critical density above which dust affects the size of a Str\"{o}mgren sphere is $n_{\rm crit} \sim 100\: Q_{49}^{-1} T_{4}^{-0.84} \sigma_{ext,-21}^{-3}$ (see paper I), where $Q_{49}$ is the ionising photon luminosity in $10^{49}$ photons s$^{-1}$, $T_4 = T/10^4 K$ is the temperature of the ionised sphere and $\sigma_{ext,-21} = \sigma_{ext}/10^{-21}$ is the extinction cross section of the dust particles. Given the standard values for our SN shock and its precursor during the rapid cooling phases (see sec \ref{subsec:rad-trans-shell}), we estimate this value to be $n_{\rm crit} \approx 100 \pcc$. We, therefore, do not expect any change in the precursor region due to dust. The formation of a dense shell ($n_H \sim 100 \pcc$) may seem to affect the radiative transfer through the shell.
However, by the time such a dense shell forms, the \HI fraction also increases inside the shell so that the \HI opacity dominates over the dust extinction. We, therefore, do not expect too much change in the radiative transfer through the dense shell either. We assess that the effect of dust may be important in the SN remnants exploding inside 
denser media ($n_H \gtrsim 100 \pcc$).

\subsection{Progenitor O/B star}
Another possible limitation of our current calculation is the absence of any pre-perturbed medium (e.g. perturbations by the progenitor star) in which the SN explodes. For example, for a progenitor star of class O7-III (mass $= 47.4 \msun$), the mass loss rate ($\dot{M}_\star$) and \HI ionising photon luminosities are $\approx 3\times 10^{-6} \mpy$ and $2\times 10^{49}$ photons s$^{-1}$, respectively for the initial $3$ Myr before the SN goes off \citep{Sternberg2003a}. Given such a large UV luminosity, the star is expected to set up a Str\"{o}mgen's sphere out to a radius of $\approx 86$ pc (for $n_H = 1 \pcc$) within the first $t_{\rm rec} \sim 1/n_H \alpha(T) \sim 120$ kyr of its evolution.
In this context, it is interesting to note that while many core-collapse SNe indeed likely completely ionise their circumstellar medium before exploding or as they explode, SNe Ia seem to explode into at least partially neutral material, as is evident by their Ly$\alpha$ emission (e.g. \citealt{Ghavamian2007}).
 While the ionising radiation  
is not expected to change the density inside the ionised sphere by much (see paper I), the stellar wind from the star can change the density distribution to a large extent. The mechanical luminosity injected by such a star is $\frac{1}{2}\: \dot{M}_\star\:v_w^2 \approx 5.5 \times 10^{36} \ergps$, considering the wind velocity $ v_w \approx 2400\: \kmps$ \citep{Sternberg2003a}. The corresponding bubble will sweep out the gas to a thin shell at a distance of $\approx 63$ pc after $3$ Myr (when the star presumably goes off as SN) of its evolution inside a $n_H = 1 \pcc$ medium \citep{Castor1975, Weaver1977}. This is clearly, larger than the shock radius from the SN within first $500$ kyr. Therefore, the SN will explode in a region with density much lower than $1 \pcc$ and the SN energy will be quickly re-distributed to the already existing shell created by the stellar wind. The evolution of the SN shock across the shell is complex  given that the shock now travels through an already collapsed high density and non-uniform shell whose density structure can be only found via simulations. 

\section{summary}
\label{sec:summary}
We have performed simulations of a spherically symmetric expanding SN remnant up to $500$ kyr into an initially uniform background medium with hydrogen density, $n_0 = 1 \pcc$. We consider non-equilibrium ionisation network, conduction, frequency dependent radiative transport and simple dust effects in addition to the usual hydrodynamics. 
The self-consistent treatment of these processes has lead us to obtain very detailed understanding of the SN remnant evolution, many aspects for the first time. Our understanding of the remnant evolution can be summarised as follows:
\begin{itemize}
	\item We find that the presence of complex physics, like the dynamically evolving  ion network, conduction and radiative transport does not alter the dynamics or energetics of the remnant compared to a simple model in which collisional ionisation equilibrium is assumed. Therefore, the thermal energy and momentum feedback from the SN to the ISM is not significantly affected  
	by the inclusion of more detailed processes 
	including the complex physics of non-equilibrium ionisation, radiative transfer or conduction (ref Fig \ref{fig:energetics}). 
	
	\item The cooling function of the material behind the shock (down to the outer radius of the hot bubble) does not follow any particular known cooling function throughout its evolution. Rather, it goes through a mixture of them. At earlier times ($t < t_{\rm cool, onset}$), the cooling follows simple CIE cooling due to high enough temperature ($T \gtrsim 10^6$ K) at the shock. During the rapid cooling phase ($t_{\rm cool, onset} \lesssim t \lesssim t_{\rm cool, end}$), the cooling function follows a simple isochoric cooling of a photo-ionised plasma due to the presence of a large  
	number of \HI ionising photons at this phase. 
	In the snow-plow phase ($t \gtrsim t_{\rm cool, end}$), the cooling function is enhanced by more than an order of magnitude compared to a simple CIE model due to the presence of different ions that are  
	out of equilibrium (ref Fig \ref{fig:cool-curve}).
	
	\item The cooling shell during the rapid cooling phase, can be as bright as an O star in  terms of its UV luminosity but only for $\approx 20$ kyr. This sets a precursor region up to $\sim 10$ pc ahead of the shock (for $n_H = 1 \pcc$). This precursor region does not expand further with time but the ions inside this region recombine to produce a precursor region that decreases with time. The ionisation level inside the precursor region is, therefore, not a direct function of the velocity as in the steady state, plane-parallel shock models, rather it depends on the recombination time of the ions. Although the precursor region persists ahead of the shock until $v_s \lesssim 70 \kmps$, consistent with the plane-parallel, steady state shock, the reason behind it is  different (ref Fig \ref{fig:vel-Fuv}).
	
	\item The presence of ions like \civ, \nv, \ovi etc in excess of $10^{14} \pcmsq$ (for $n_0 = 1.0 \pcc$) implies that the SN remnant is undergoing a rapid cooling phase. These ions are expected to remain over-ionised inside the bubble due to delayed recombination.
	
	\item The observable column densities for different ions can differ by a factor between
	$5$ and a few orders of magnitude compared to estimations from plane-parallel steady state shocks (Fig \ref{fig:all-colden}). Major differences are i) the non-steady behaviour of the shock, ii) the geometrical factors and iii) the non-steady precursor. Such fundamental drawbacks of the steady state shock models lead to severe underestimation the presence of many ions and should be used cautiously.
	
\end{itemize}

In conclusion, despite several limitations of the simulations performed in this paper, we believe that we have been able to understand some fundamental aspects of the radiative transport and non-equilibrium ion dynamics in a SN remnant and a spherical shock, in general. Challenges, however, remain to deal with the limitations of spherical symmetry, magnetic field and initial density distribution due to the progenitor star. We hope to address them in our future works.

\section*{acknowledgements}
We acknowledge the helpful discussion by Yakov Faerman and Eve Ostriker. We thank the Center for Computational Astrophysics (CCA) at the Flatiron Institute Simons Foundation for hospitality and for computational support via the Scientific Computing Core. We also thank the anonymous referee for his/her comments that improved the content of this paper. This work was supported by the German Science Foundation via DFG/DIP grant STE 1869/2-1 GE625/17-1 at Tel-Aviv University, the Israeli Centers of Excellence (I-CORE) program (center no. 1829/12) and the Israeli Science Foundation (ISF grant no. 2190/20).

\section*{Data availability}
The data underlying this article are available in the article and in its online supplementary material.

 
\bibliography{library}

\begin{thebibliography}{}
\makeatletter
\relax
\def\mn@urlcharsother{\let\do\@makeother \do\$\do\&\do\#\do\^\do\_\do\%\do\~}
\def\mn@doi{\begingroup\mn@urlcharsother \@ifnextchar [ {\mn@doi@}
  {\mn@doi@[]}}
\def\mn@doi@[#1]#2{\def\@tempa{#1}\ifx\@tempa\@empty \href
  {http://dx.doi.org/#2} {doi:#2}\else \href {http://dx.doi.org/#2} {#1}\fi
  \endgroup}
\def\mn@eprint#1#2{\mn@eprint@#1:#2::\@nil}
\def\mn@eprint@arXiv#1{\href {http://arxiv.org/abs/#1} {{\tt arXiv:#1}}}
\def\mn@eprint@dblp#1{\href {http://dblp.uni-trier.de/rec/bibtex/#1.xml}
  {dblp:#1}}
\def\mn@eprint@#1:#2:#3:#4\@nil{\def\@tempa {#1}\def\@tempb {#2}\def\@tempc
  {#3}\ifx \@tempc \@empty \let \@tempc \@tempb \let \@tempb \@tempa \fi \ifx
  \@tempb \@empty \def\@tempb {arXiv}\fi \@ifundefined
  {mn@eprint@\@tempb}{\@tempb:\@tempc}{\expandafter \expandafter \csname
  mn@eprint@\@tempb\endcsname \expandafter{\@tempc}}}

\bibitem[\protect\citeauthoryear{Bach}{Bach}{2019}]{Bach2019}
Bach J.~B.,  2019, PhD thesis

\bibitem[\protect\citeauthoryear{Becker, Holt, Smith, White  \& Boldt}{Becker
  et~al.}{1980}]{Becker1980a}
Becker R.~H.,  Holt S.~S.,  Smith B.~W.,  White N.~E.,   Boldt E.~A.,  1980,
  \mn@doi [ApJ] {10.3130/aijs.65.65_1}, 235, 5

\bibitem[\protect\citeauthoryear{Breitschwerdt \& de Avillez}{Breitschwerdt \&
  de~Avillez}{2006}]{Breitschwerdt2006}
Breitschwerdt D.,  de Avillez M.~A.,  2006, \mn@doi [Astron. Astrophys.]
  {10.1051/0004-6361:20064989}, 452, L1

\bibitem[\protect\citeauthoryear{Castor, Weaver  \& McCray}{Castor
  et~al.}{1975}]{Castor1975}
Castor J.,  Weaver R.,   McCray R.,  1975, \mn@doi [Astrophys. J.]
  {10.1086/181908}, 200, L107

\bibitem[\protect\citeauthoryear{{Chevalier} \& {Imamura}}{{Chevalier} \&
  {Imamura}}{1982}]{Chevalier1982}
{Chevalier} R.~A.,  {Imamura} J.~N.,  1982, \mn@doi [\apj] {10.1086/160364},
  \href {https://ui.adsabs.harvard.edu/abs/1982ApJ...261..543C} {261, 543}

\bibitem[\protect\citeauthoryear{{Cioffi}, {McKee}  \& {Bertschinger}}{{Cioffi}
  et~al.}{1988}]{Cioffi1988}
{Cioffi} D.~F.,  {McKee} C.~F.,   {Bertschinger} E.,  1988, \mn@doi [\apj]
  {10.1086/166834}, \href
  {https://ui.adsabs.harvard.edu/abs/1988ApJ...334..252C} {334, 252}

\bibitem[\protect\citeauthoryear{Cowie \& McKee}{Cowie \&
  McKee}{1977}]{Cowie1977}
Cowie L.~L.,  McKee C.~F.,  1977, \mn@doi [Astrophys. J.] {10.1086/154911},
  211, 135

\bibitem[\protect\citeauthoryear{Cox}{Cox}{1972a}]{Cox1972}
Cox D.~P.,  1972a, \mn@doi [ApJ] {10.1086/151774}, \href
  {https://ui.adsabs.harvard.edu/abs/1972ApJ...178..143C} {178, 143}

\bibitem[\protect\citeauthoryear{{Cox}}{{Cox}}{1972b}]{Cox1972a}
{Cox} D.~P.,  1972b, \mn@doi [\apj] {10.1086/151775}, \href
  {https://ui.adsabs.harvard.edu/abs/1972ApJ...178..159C} {178, 159}

\bibitem[\protect\citeauthoryear{{Cox}}{{Cox}}{2005}]{Cox2005}
{Cox} D.~P.,  2005, \mn@doi [\araa] {10.1146/annurev.astro.43.072103.150615},
  \href {https://ui.adsabs.harvard.edu/abs/2005ARA&A..43..337C} {43, 337}

\bibitem[\protect\citeauthoryear{{Cox} \& {Anderson}}{{Cox} \&
  {Anderson}}{1982}]{Cox1982}
{Cox} D.~P.,  {Anderson} P.~R.,  1982, \mn@doi [\apj] {10.1086/159632}, \href
  {https://ui.adsabs.harvard.edu/abs/1982ApJ...253..268C} {253, 268}

\bibitem[\protect\citeauthoryear{Cui \& Cox}{Cui \& Cox}{1992}]{Cui1992}
Cui W.,  Cox D.~P.,  1992, ApJ, 401, 206

\bibitem[\protect\citeauthoryear{{De Avillez} \& Breitschwerdt}{{De Avillez} \&
  Breitschwerdt}{2012}]{DeAvillez2012}
{De Avillez} M.~A.,  Breitschwerdt D.,  2012, \mn@doi [Astrophys. J. Lett.]
  {10.1088/2041-8205/761/2/L19}, 761, 6

\bibitem[\protect\citeauthoryear{Dekel \& Silk}{Dekel \&
  Silk}{1986}]{Dekel1986}
Dekel A.,  Silk J.,  1986, \mn@doi [Astrophys. J.] {10.1086/164050}, 303, 39

\bibitem[\protect\citeauthoryear{{Dekel}, {Sarkar}, {Jiang}, {Bournaud},
  {Krumholz}, {Ceverino}  \& {Primack}}{{Dekel} et~al.}{2019}]{Dekel2019}
{Dekel} A.,  {Sarkar} K.~C.,  {Jiang} F.,  {Bournaud} F.,  {Krumholz} M.~R.,
  {Ceverino} D.,   {Primack} J.~R.,  2019, \mn@doi [\mnras]
  {10.1093/mnras/stz1919}, \href
  {https://ui.adsabs.harvard.edu/abs/2019MNRAS.488.4753D} {488, 4753}

\bibitem[\protect\citeauthoryear{{Dopita}}{{Dopita}}{1976}]{Dopita1976}
{Dopita} M.~A.,  1976, \mn@doi [\apj] {10.1086/154732}, \href
  {https://ui.adsabs.harvard.edu/abs/1976ApJ...209..395D} {209, 395}

\bibitem[\protect\citeauthoryear{{Dopita} \& {Sutherland}}{{Dopita} \&
  {Sutherland}}{1996}]{Dopita1996}
{Dopita} M.~A.,  {Sutherland} R.~S.,  1996, \mn@doi [\apjs] {10.1086/192255},
  \href {https://ui.adsabs.harvard.edu/abs/1996ApJS..102..161D} {102, 161}

\bibitem[\protect\citeauthoryear{{Dopita}, {Dodorico}  \& {Benvenuti}}{{Dopita}
  et~al.}{1980}]{Dopita1980}
{Dopita} M.~A.,  {Dodorico} S.,   {Benvenuti} P.,  1980, \mn@doi [\apj]
  {10.1086/157781}, \href
  {https://ui.adsabs.harvard.edu/abs/1980ApJ...236..628D} {236, 628}

\bibitem[\protect\citeauthoryear{Dopita, Seitenzahl, Sutherland, Nicholls,
  Vogt, Ghavamian  \& Ruiter}{Dopita et~al.}{2019}]{Dopita2019}
Dopita M.~A.,  Seitenzahl I.~R.,  Sutherland R.~S.,  Nicholls D.~C.,  Vogt F.
  P.~A.,  Ghavamian P.,   Ruiter A.~J.,  2019, \mn@doi [The Astronomical
  Journal] {10.3847/1538-3881/aaf235}, 157, 50

\bibitem[\protect\citeauthoryear{{Draine}}{{Draine}}{2011}]{Draine2011}
{Draine} B.~T.,  2011, {Physics of the Interstellar and Intergalactic Medium}.
Princeton University Press

\bibitem[\protect\citeauthoryear{{Duffell}}{{Duffell}}{2016}]{Duffell2016}
{Duffell} P.~C.,  2016, \mn@doi [\apj] {10.3847/0004-637X/821/2/76}, \href
  {https://ui.adsabs.harvard.edu/abs/2016ApJ...821...76D} {821, 76}

\bibitem[\protect\citeauthoryear{{Edmiston} \& {Kennel}}{{Edmiston} \&
  {Kennel}}{1984}]{Ediston1984}
{Edmiston} J.~P.,  {Kennel} C.~F.,  1984, \mn@doi [Journal of Plasma Physics]
  {10.1017/S002237780000218X}, \href
  {https://ui.adsabs.harvard.edu/abs/1984JPlPh..32..429E} {32, 429}

\bibitem[\protect\citeauthoryear{{El-Badry}, {Ostriker}, {Kim}, {Quataert}  \&
  {Weisz}}{{El-Badry} et~al.}{2019}]{ElBadry2019}
{El-Badry} K.,  {Ostriker} E.~C.,  {Kim} C.-G.,  {Quataert} E.,   {Weisz}
  D.~R.,  2019, \mn@doi [\mnras] {10.1093/mnras/stz2773}, \href
  {https://ui.adsabs.harvard.edu/abs/2019MNRAS.490.1961E} {490, 1961}

\bibitem[\protect\citeauthoryear{{Evirgen} \& {Gent}}{{Evirgen} \&
  {Gent}}{2019}]{Evirgen2019}
{Evirgen} C.~C.,  {Gent} F.,  2019, arXiv e-prints, \href
  {https://ui.adsabs.harvard.edu/abs/2019arXiv190808781E} {p. arXiv:1908.08781}

\bibitem[\protect\citeauthoryear{{Fesen} \& {Kirshner}}{{Fesen} \&
  {Kirshner}}{1980}]{Fesen1980}
{Fesen} R.~A.,  {Kirshner} R.~P.,  1980, \mn@doi [\apj] {10.1086/158534}, \href
  {https://ui.adsabs.harvard.edu/abs/1980ApJ...242.1023F} {242, 1023}

\bibitem[\protect\citeauthoryear{{Fesen}, {Blair}  \& {Kirshner}}{{Fesen}
  et~al.}{1985}]{Fesen1985}
{Fesen} R.~A.,  {Blair} W.~P.,   {Kirshner} R.~P.,  1985, \mn@doi [\apj]
  {10.1086/163130}, \href
  {https://ui.adsabs.harvard.edu/abs/1985ApJ...292...29F} {292, 29}

\bibitem[\protect\citeauthoryear{{Ghavamian}, {Laming}  \&
  {Rakowski}}{{Ghavamian} et~al.}{2007}]{Ghavamian2007}
{Ghavamian} P.,  {Laming} J.~M.,   {Rakowski} C.~E.,  2007, \mn@doi [\apjl]
  {10.1086/510740}, \href
  {https://ui.adsabs.harvard.edu/abs/2007ApJ...654L..69G} {654, L69}

\bibitem[\protect\citeauthoryear{Gnat}{Gnat}{2017}]{Gnat2017}
Gnat O.,  2017, \mn@doi [Astrophys. J. Suppl. Ser.]
  {10.3847/1538-4365/228/2/11}, 228, 1

\bibitem[\protect\citeauthoryear{Gnat \& Sternberg}{Gnat \&
  Sternberg}{2007}]{Gnat2007}
Gnat O.,  Sternberg A.,  2007, \mn@doi [ApJSS] {10.1086/509786}, 168, 213

\bibitem[\protect\citeauthoryear{{Gnat} \& {Sternberg}}{{Gnat} \&
  {Sternberg}}{2009}]{Gnat2009}
{Gnat} O.,  {Sternberg} A.,  2009, \mn@doi [\apj]
  {10.1088/0004-637X/693/2/1514}, \href
  {https://ui.adsabs.harvard.edu/abs/2009ApJ...693.1514G} {693, 1514}

\bibitem[\protect\citeauthoryear{{Grun}, {Stamper}, {Manka}, {Resnick},
  {Burris}, {Crawford}  \& {Ripin}}{{Grun} et~al.}{1991}]{Grun1991}
{Grun} J.,  {Stamper} J.,  {Manka} C.,  {Resnick} J.,  {Burris} R.,  {Crawford}
  J.,   {Ripin} B.~H.,  1991, \mn@doi [\prl] {10.1103/PhysRevLett.66.2738},
  \href {https://ui.adsabs.harvard.edu/abs/1991PhRvL..66.2738G} {66, 2738}

\bibitem[\protect\citeauthoryear{Haardt \& Madau}{Haardt \&
  Madau}{2012}]{Haardt2012}
Haardt F.,  Madau P.,  2012, \mn@doi [Astrophys. J.]
  {10.1088/0004-637X/746/2/125}, 746

\bibitem[\protect\citeauthoryear{{Hamilton}, {Sarazin}  \&
  {Chevalier}}{{Hamilton} et~al.}{1983}]{Hamilton1983}
{Hamilton} A.~J.~S.,  {Sarazin} C.~L.,   {Chevalier} R.~A.,  1983, \mn@doi
  [\apjs] {10.1086/190841}, \href
  {https://ui.adsabs.harvard.edu/abs/1983ApJS...51..115H} {51, 115}

\bibitem[\protect\citeauthoryear{{Itoh}}{{Itoh}}{1978}]{Itoh1978}
{Itoh} H.,  1978, \pasj, \href
  {https://ui.adsabs.harvard.edu/abs/1978PASJ...30..489I} {30, 489}

\bibitem[\protect\citeauthoryear{{Kafatos}}{{Kafatos}}{1973}]{Kafatos1973}
{Kafatos} M.,  1973, \mn@doi [\apj] {10.1086/152151}, \href
  {https://ui.adsabs.harvard.edu/abs/1973ApJ...182..433K} {182, 433}

\bibitem[\protect\citeauthoryear{{Kewley}, {Dopita}, {Sutherland}, {Heisler}
  \& {Trevena}}{{Kewley} et~al.}{2001}]{Kewley2001}
{Kewley} L.~J.,  {Dopita} M.~A.,  {Sutherland} R.~S.,  {Heisler} C.~A.,
  {Trevena} J.,  2001, \mn@doi [\apj] {10.1086/321545}, \href
  {https://ui.adsabs.harvard.edu/abs/2001ApJ...556..121K} {556, 121}

\bibitem[\protect\citeauthoryear{Kim \& Ostriker}{Kim \&
  Ostriker}{2015}]{Kim2015}
Kim C.~G.,  Ostriker E.~C.,  2015, \mn@doi [Astrophys. J.]
  {10.1088/0004-637X/802/2/99}, 802, 1

\bibitem[\protect\citeauthoryear{{Kopsacheili}, {Zezas}  \&
  {Leonidaki}}{{Kopsacheili} et~al.}{2020}]{Kopsacheili2020}
{Kopsacheili} M.,  {Zezas} A.,   {Leonidaki} I.,  2020, \mn@doi [\mnras]
  {10.1093/mnras/stz2594}, \href
  {https://ui.adsabs.harvard.edu/abs/2020MNRAS.491..889K} {491, 889}

\bibitem[\protect\citeauthoryear{{Krause}, {Fierlinger}, {Diehl}, {Burkert},
  {Voss}  \& {Ziegler}}{{Krause} et~al.}{2013}]{Krause2013}
{Krause} M.,  {Fierlinger} K.,  {Diehl} R.,  {Burkert} A.,  {Voss} R.,
  {Ziegler} U.,  2013, \mn@doi [\aap] {10.1051/0004-6361/201220060}, \href
  {https://ui.adsabs.harvard.edu/abs/2013A&A...550A..49K} {550, A49}

\bibitem[\protect\citeauthoryear{Krumholz., Kruijssen  \& Crocker}{Krumholz.
  et~al.}{2017}]{Krumholz2016}
Krumholz. M.~R.,  Kruijssen J. M.~D.,   Crocker R.~M.,  2017, \mn@doi [Mon.
  Not. R. Astron. Soc] {10.1093/mnras/stw3195}, 466, 1213

\bibitem[\protect\citeauthoryear{Krumholz, Burkhart, Forbes  \&
  Crocker}{Krumholz et~al.}{2018}]{Krumholz2018}
Krumholz M.~R.,  Burkhart B.,  Forbes J.~C.,   Crocker R.~M.,  2018, \mn@doi
  [Mon. Not. R. Astron. Soc.] {10.1093/MNRAS/STY852}, 477, 2716

\bibitem[\protect\citeauthoryear{{Laming}}{{Laming}}{2001}]{Laming2001}
{Laming} J.~M.,  2001, \mn@doi [\apj] {10.1086/318317}, \href
  {https://ui.adsabs.harvard.edu/abs/2001ApJ...546.1149L} {546, 1149}

\bibitem[\protect\citeauthoryear{{Laming}}{{Laming}}{2004}]{Laming2004}
{Laming} J.~M.,  2004, \mn@doi [\pre] {10.1103/PhysRevE.70.057402}, \href
  {https://ui.adsabs.harvard.edu/abs/2004PhRvE..70e7402L} {70, 057402}

\bibitem[\protect\citeauthoryear{{Laming} \& {Grun}}{{Laming} \&
  {Grun}}{2002}]{Laming2002}
{Laming} J.~M.,  {Grun} J.,  2002, \mn@doi [\prl]
  {10.1103/PhysRevLett.89.125002}, \href
  {https://ui.adsabs.harvard.edu/abs/2002PhRvL..89l5002L} {89, 125002}

\bibitem[\protect\citeauthoryear{{Laming} \& {Grun}}{{Laming} \&
  {Grun}}{2003}]{Laming2003}
{Laming} J.~M.,  {Grun} J.,  2003, \mn@doi [Physics of Plasmas]
  {10.1063/1.1556603}, \href
  {https://ui.adsabs.harvard.edu/abs/2003PhPl...10.1614L} {10, 1614}

\bibitem[\protect\citeauthoryear{Larson}{Larson}{1974}]{Larson1974}
Larson R.~B.,  1974, \mn@doi [Mon. Not. R. Astron. Soc]
  {10.1093/mnras/169.2.229}, 169, 229

\bibitem[\protect\citeauthoryear{{Mathewson} \& {Clarke}}{{Mathewson} \&
  {Clarke}}{1973}]{Mathewson1973}
{Mathewson} D.~S.,  {Clarke} J.~N.,  1973, \mn@doi [\apj] {10.1086/152002},
  \href {https://ui.adsabs.harvard.edu/abs/1973ApJ...180..725M} {180, 725}

\bibitem[\protect\citeauthoryear{{McKee} \& {Ostriker}}{{McKee} \&
  {Ostriker}}{1977}]{McKee1977}
{McKee} C.~F.,  {Ostriker} J.~P.,  1977, \mn@doi [\apj] {10.1086/155667}, \href
  {https://ui.adsabs.harvard.edu/abs/1977ApJ...218..148M} {218, 148}

\bibitem[\protect\citeauthoryear{Mignone, Bodo, Massaglia, Matsakos, Tesileanu,
  Zanni  \& Ferrari}{Mignone et~al.}{2007}]{Mignone2007}
Mignone A.,  Bodo G.,  Massaglia S.,  Matsakos T.,  Tesileanu O.,  Zanni C.,
  Ferrari A.,  2007, \mn@doi [Astrophys. J. Suppl. Ser.] {10.1086/513316}, 170,
  228

\bibitem[\protect\citeauthoryear{Nath \& Trentham}{Nath \&
  Trentham}{1997}]{Nath1997}
Nath B.~B.,  Trentham N.,  1997, Mon. Not. R. Astron. Soc, 291, 505

\bibitem[\protect\citeauthoryear{{Nussbaumer} \& {Schmutz}}{{Nussbaumer} \&
  {Schmutz}}{1984}]{Nussbaumer1984}
{Nussbaumer} H.,  {Schmutz} W.,  1984, \aap, \href
  {https://ui.adsabs.harvard.edu/abs/1984A&A...138..495N} {138, 495}

\bibitem[\protect\citeauthoryear{{Okon} et~al.,}{{Okon}
  et~al.}{2019}]{Okon2019}
{Okon} H.,  et~al., 2019, arXiv e-prints, \href
  {https://ui.adsabs.harvard.edu/abs/2019arXiv191208129O} {p. arXiv:1912.08129}

\bibitem[\protect\citeauthoryear{{Petruk}, {Kuzyo}  \& {Beshley}}{{Petruk}
  et~al.}{2016}]{Patruk2016}
{Petruk} O.,  {Kuzyo} T.,   {Beshley} V.,  2016, \mn@doi [\mnras]
  {10.1093/mnras/stv2746}, \href
  {https://ui.adsabs.harvard.edu/abs/2016MNRAS.456.2343P} {456, 2343}

\bibitem[\protect\citeauthoryear{{Pittard}, {Dobson}, {Durisen}, {Dyson},
  {Hartquist}  \& {O'Brien}}{{Pittard} et~al.}{2005}]{Pittard2005}
{Pittard} J.~M.,  {Dobson} M.~S.,  {Durisen} R.~H.,  {Dyson} J.~E.,
  {Hartquist} T.~W.,   {O'Brien} J.~T.,  2005, \mn@doi [\aap]
  {10.1051/0004-6361:20042260}, \href
  {https://ui.adsabs.harvard.edu/abs/2005A&A...438...11P} {438, 11}

\bibitem[\protect\citeauthoryear{{Quest}}{{Quest}}{1988}]{Quest1988}
{Quest} K.~B.,  1988, \mn@doi [\jgr] {10.1029/JA093iA09p09649}, \href
  {https://ui.adsabs.harvard.edu/abs/1988JGR....93.9649Q} {93, 9649}

\bibitem[\protect\citeauthoryear{{Ritchey}}{{Ritchey}}{2020}]{Ritchey2020b}
{Ritchey} A.~M.,  2020, \mn@doi [\mnras] {10.1093/mnras/staa1375}, \href
  {https://ui.adsabs.harvard.edu/abs/2020MNRAS.495.2909R} {495, 2909}

\bibitem[\protect\citeauthoryear{{Ritchey}, {Jenkins}, {Federman}, {Rice},
  {Caprioli}  \& {Wallerstein}}{{Ritchey} et~al.}{2020}]{Ritchey2020a}
{Ritchey} A.~M.,  {Jenkins} E.~B.,  {Federman} S.~R.,  {Rice} J.~S.,
  {Caprioli} D.,   {Wallerstein} G.,  2020, arXiv e-prints, \href
  {https://ui.adsabs.harvard.edu/abs/2020arXiv200509096R} {p. arXiv:2005.09096}

\bibitem[\protect\citeauthoryear{{Sarkar}, {Sternberg}  \& {Gnat}}{{Sarkar}
  et~al.}{2020}]{Sarkar2020a}
{Sarkar} K.~C.,  {Sternberg} A.,   {Gnat} O.,  2020, arXiv e-prints, \href
  {https://ui.adsabs.harvard.edu/abs/2020arXiv201000457S} {p. arXiv:2010.00457}

\bibitem[\protect\citeauthoryear{{Shull} \& {McKee}}{{Shull} \&
  {McKee}}{1979}]{Shull1979}
{Shull} J.~M.,  {McKee} C.~F.,  1979, \mn@doi [\apj] {10.1086/156712}, \href
  {https://ui.adsabs.harvard.edu/abs/1979ApJ...227..131S} {227, 131}

\bibitem[\protect\citeauthoryear{Slavin \& Cox}{Slavin \&
  Cox}{1992}]{Slavin1992}
Slavin J.~D.,  Cox D.~P.,  1992, ApJ, 392, 131

\bibitem[\protect\citeauthoryear{{Spitzer}}{{Spitzer}}{1956}]{Spitzer1956}
{Spitzer} L.,  1956, {Physics of Fully Ionized Gases}.
Dover Publications

\bibitem[\protect\citeauthoryear{{Steinberg} \& {Metzger}}{{Steinberg} \&
  {Metzger}}{2018}]{Steinberg2018}
{Steinberg} E.,  {Metzger} B.~D.,  2018, \mn@doi [\mnras]
  {10.1093/mnras/sty1641}, \href
  {https://ui.adsabs.harvard.edu/abs/2018MNRAS.479..687S} {479, 687}

\bibitem[\protect\citeauthoryear{{Steinwandel}, {Moster}, {Naab}, {Hu}  \&
  {Walch}}{{Steinwandel} et~al.}{2020}]{Steinwandel2020}
{Steinwandel} U.~P.,  {Moster} B.~P.,  {Naab} T.,  {Hu} C.-Y.,   {Walch} S.,
  2020, \mn@doi [\mnras] {10.1093/mnras/staa821}, \href
  {https://ui.adsabs.harvard.edu/abs/2020MNRAS.495.1035S} {495, 1035}

\bibitem[\protect\citeauthoryear{Sternberg, Hoffmann  \& Pauldrach}{Sternberg
  et~al.}{2003}]{Sternberg2003a}
Sternberg A.,  Hoffmann T.~L.,   Pauldrach A. W.~A.,  2003, \mn@doi [Astrophys.
  J.] {10.1086/379506}, 599, 1333

\bibitem[\protect\citeauthoryear{{Sutherland} \& {Dopita}}{{Sutherland} \&
  {Dopita}}{2017}]{Sutherland2017}
{Sutherland} R.~S.,  {Dopita} M.~A.,  2017, \mn@doi [\apjs]
  {10.3847/1538-4365/aa6541}, \href
  {https://ui.adsabs.harvard.edu/abs/2017ApJS..229...34S} {229, 34}

\bibitem[\protect\citeauthoryear{Teşileanu, Mignone  \& Massaglia}{Teşileanu
  et~al.}{2008}]{Tesileanu2008}
Teşileanu O.,  Mignone A.,   Massaglia S.,  2008, \mn@doi [Astron. Astrophys.]
  {10.1051/0004-6361:200809461}, 488, 429

\bibitem[\protect\citeauthoryear{{Treumann}}{{Treumann}}{2009}]{Treumann2009}
{Treumann} R.~A.,  2009, \mn@doi [\aapr] {10.1007/s00159-009-0024-2}, \href
  {https://ui.adsabs.harvard.edu/abs/2009A&ARv..17..409T} {17, 409}

\bibitem[\protect\citeauthoryear{Vasiliev}{Vasiliev}{2013}]{Vasiliev2013a}
Vasiliev E.~O.,  2013, \mn@doi [Mon. Not. R. Astron. Soc.]
  {10.1093/mnras/stt189}, 431, 638

\bibitem[\protect\citeauthoryear{{Vishniac}}{{Vishniac}}{1983}]{Vishniac1983}
{Vishniac} E.~T.,  1983, \mn@doi [\apj] {10.1086/161433}, \href
  {https://ui.adsabs.harvard.edu/abs/1983ApJ...274..152V} {274, 152}

\bibitem[\protect\citeauthoryear{Weaver, McCray, Castor, Shapiro  \&
  Moore}{Weaver et~al.}{1977}]{Weaver1977}
Weaver R.,  McCray R.,  Castor J.,  Shapiro P.,   Moore R.,  1977, \mn@doi
  [Astrophys. J.] {10.1086/155692}, 218, 377

\bibitem[\protect\citeauthoryear{Weingartner \& Draine}{Weingartner \&
  Draine}{2001}]{Weingartner2001}
Weingartner J.~C.,  Draine B.~T.,  2001, \mn@doi [Astrophys. J.]
  {10.1086/318651}, 548, 296

\bibitem[\protect\citeauthoryear{Yadav, Mukherjee, Sharma, Nath, Mukherjee,
  Sharma  \& Nath}{Yadav et~al.}{2017}]{Yadav2017}
Yadav N.,  Mukherjee D.,  Sharma P.,  Nath B.~B.,  Mukherjee D.,  Sharma P.,
  Nath B.~B.,  2017, \mn@doi [Mon. Not. R. Astron. Soc]
  {10.1093/mnras/stw2522}, 465, 1720

\bibitem[\protect\citeauthoryear{{Zhang}, {Foster}  \& {Smith}}{{Zhang}
  et~al.}{2018}]{Zhang2018}
{Zhang} G.-Y.,  {Foster} A.,   {Smith} R.,  2018, \mn@doi [\apj]
  {10.3847/1538-4357/aad692}, \href
  {https://ui.adsabs.harvard.edu/abs/2018ApJ...864...79Z} {864, 79}

\bibitem[\protect\citeauthoryear{{Zhang}, {Slavin}, {Foster}, {Smith},
  {ZuHone}, {Zhou}  \& {Chen}}{{Zhang} et~al.}{2019}]{Zhang2019}
{Zhang} G.-Y.,  {Slavin} J.~D.,  {Foster} A.,  {Smith} R.~K.,  {ZuHone} J.~A.,
  {Zhou} P.,   {Chen} Y.,  2019, \mn@doi [\apj] {10.3847/1538-4357/ab0f9a},
  \href {https://ui.adsabs.harvard.edu/abs/2019ApJ...875...81Z} {875, 81}

\bibitem[\protect\citeauthoryear{{de Avillez} \& {Mac Low}}{{de Avillez} \&
  {Mac Low}}{2002}]{deAvillez2002}
{de Avillez} M.~A.,  {Mac Low} M.-M.,  2002, \mn@doi [\apj] {10.1086/344256},
  \href {https://ui.adsabs.harvard.edu/abs/2002ApJ...581.1047D} {581, 1047}

\makeatother
\end{thebibliography}
\bibliographystyle{mnras}

\appendix
\section{Energetics}
\begin{figure*}
	\centering
	\includegraphics[width=0.5\textwidth]{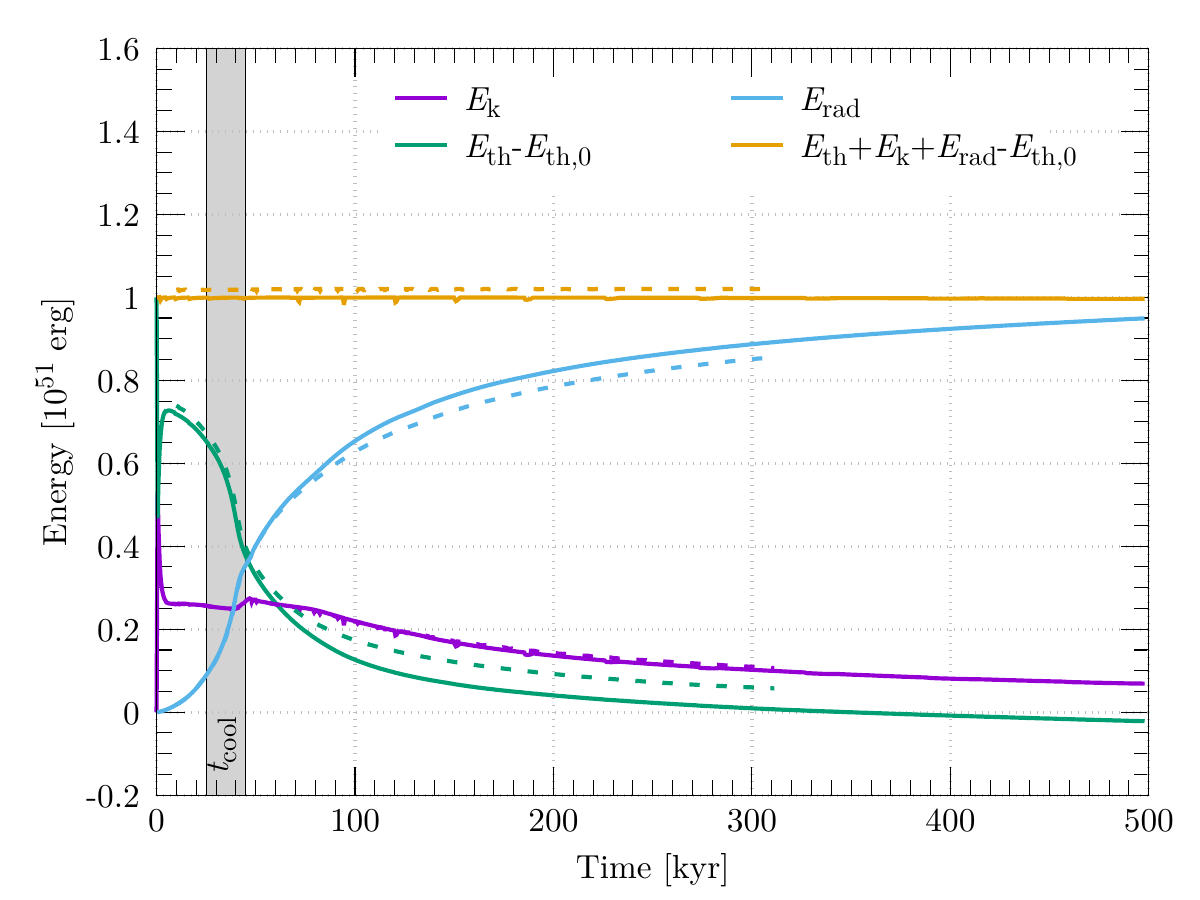}
	\caption{Evolution of energetics of a SN in a $n_H = 1 \pcc$ background medium for \nsrdc~case (dashed line). The solid line shows a simulation with resolution of $\Delta r = 0.0061$ pc and box size of $100$ pc comparing the effects of resolution on energetics. Different lines represent different energy components inside the simulation box. The green lines show the internal energy after subtracting the initial thermal energy of the box and the magenta lines show the kinetic energy. The sky-blue lines show the energy that is radiated away from the box, mostly due to the evolution of the SN remnant and a small part from the cooling of the background material. The golden lines show the total energy (kinetic + thermal + radiation loss-initial thermal energy) conservation in the simulation. 
	The extension of the rapid cooling phase is shown by the grey band near $40$ kyr. Note that the SN remnant looses almost $30$\% of its energy during this phase. The energy components flatten after $\sim 300$ kyr and rises only by a few \% after that. The reduced energy loss in high resolution simulation is due to reducing artificial cooling at the bubble-shell interface as explained in section \ref{subsubsec:convergence}. } 
	\label{app-fig:energetics}
\end{figure*}

\begin{figure*}
	\centering
	\includegraphics[width=0.8\textwidth, clip=true, trim={1.5cm 0cm 0cm 1cm}]{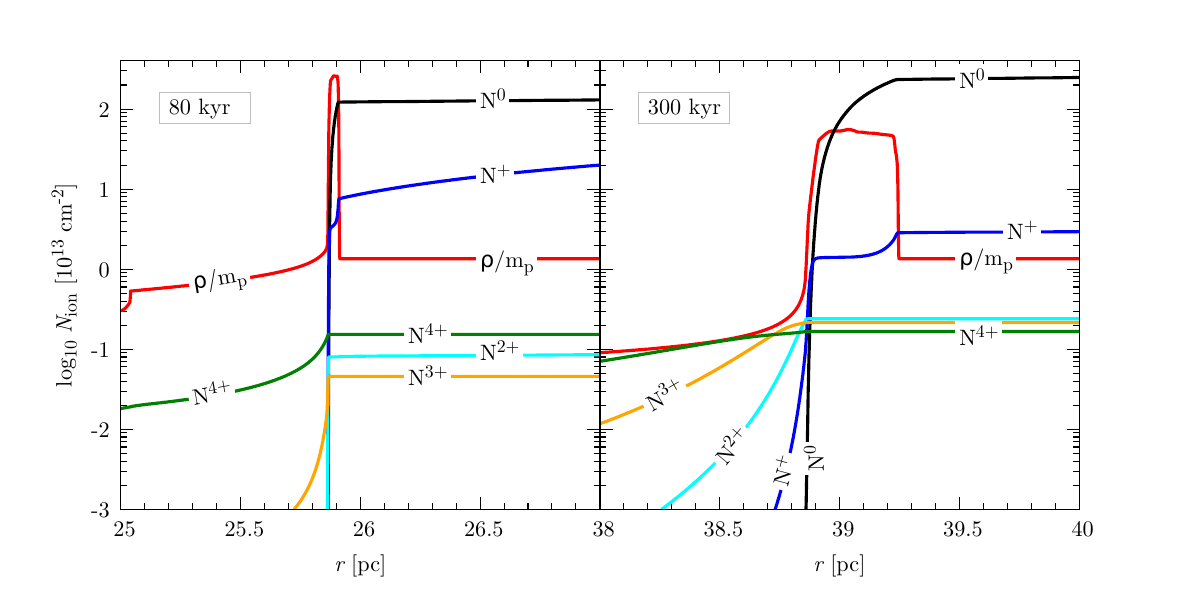}
	\caption{Origin of ions at late stages - when the precursor is present (left column) and when the precursor is not present and the bubble temperature is suitable for the production of higher ions (right column).}
	\label{fig:origin-of-ions}
\end{figure*}

\label{lastpage}
\end{document}